\documentclass{article}
\pdfoutput=1
\usepackage{arxiv}
\usepackage{xspace}
\usepackage{amsmath,amsfonts,amssymb}
\usepackage{physics}
\usepackage{graphicx}
\usepackage{algorithm}
\usepackage{algpseudocode}
\usepackage{caption}
\usepackage[makeroom]{cancel}
\usepackage{xcolor}
\usepackage[normalem]{ulem}
\usepackage[backend=biber,style=numeric,giveninits=true,
sorting=ynt]{biblatex}
\newcommand{\citep}[1]{\parencite{#1}}
\newcommand{\citet}[1]{\textcite{#1}}

\newcommand{\EMmodif}[2]{{ #2}}

\newcommand*{\mbf}[1]{\mathbf{#1}}
\newcommand{\transpose}{\mathsf{T}}
\newcommand{\ddroit}{\mathrm{d}}
\newcommand*{\eg}{\textit{e.g.}\@\xspace}

\makeatletter
\newcommand*{\etc}{%
    \@ifnextchar{.}%
        {etc}%
        {etc.\@\xspace}%
}
\makeatother
\title{CD-ROM : Complemented Deep - Reduced Order Model}

\author{Emmanuel Menier\\
	Tau / LISN / CNRS\\
	IRT SystemX\\
	Universite Paris-Saclay\\
	Orsay, France\\
	\texttt{emmanuel.menier@inria.fr}\\
	\And
	Michele Alessandro Bucci\\
	Inria / LISN / CNRS\\
	Universite Paris-Saclay\\
	Orsay, France\\
	\And
	Mouadh Yagoubi\\
	IRT SystemX\\
	Palaiseau France\\
	\And
	Lionel Mathelin\\
	LISN / CNRS\\
	Universite Paris-Saclay\\
	Orsay, France\\
	\And	
	Marc Schoenauer\\
	Inria / LISN / CNRS\\
	Universite Paris-Saclay\\
	Orsay, France
}

\renewcommand{\shorttitle}{CD-ROM : Complemented Deep - Reduced Order Model}
\begin{document}

\maketitle

\begin{abstract}
    Model order reduction through the POD-Galerkin method can lead to dramatic gains in terms of computational efficiency in solving physical problems. However, the applicability of the method to non linear high-dimensional dynamical systems such as the Navier-Stokes equations has been shown to be limited, producing inaccurate and sometimes unstable models. This paper proposes a deep learning based closure modeling approach for classical POD-Galerkin reduced order models (ROM). The proposed approach is theoretically grounded, using neural networks to approximate well studied operators. In contrast with most previous works, the present CD-ROM approach is based on an interpretable continuous memory formulation, derived from simple hypotheses on the behavior of partially observed dynamical systems. The final corrected models can hence be simulated using most classical time stepping schemes. The capabilities of the CD-ROM approach are demonstrated on two classical examples from Computational Fluid Dynamics, as well as a parametric case, the Kuramoto-Sivashinsky equation. 
\end{abstract}

\maketitle

\section{Introduction}
\subsection{General context}

The simulation of complex physical processes requires solving high dimensional partial differential equations (PDE), often at large numerical cost. This constitutes a major limitation on the applicability of accurate simulation methods to engineering problems, such as model predictive control or iterative design optimisation. Despite this complexity, physical problems can often be suitably approximated by a reduced number of dominant structures, suggesting the existence of low-dimensional manifolds supporting the underlying dynamics and fueling the development of reduction methods.

The construction of representative reduced order models often relies on the identification of a suitable modal basis to project a description of the system onto. These modes are most often extracted from data, through different methods. For example, the well-known Proper Orthogonal Decomposition (POD)-Galerkin method (\cite{lumley1970pod}) has been applied to the reduction of linear systems with great results (\cite{rowley2016review}). It was also applied to more complex problems such as the Navier Stokes equations with varying degrees of success (\cite{Berkooz2003pod,Deane1991pod}). A three dimensional Galerkin model was shown in \cite{Noack2003cylinder} to capture the transient and oscillation regimes of the cylinder flow, provided the POD basis included a mode tailored to support the transition. Models based on Galerkin POD formulations have the advantage of reducing computational cost while preserving a high level of interpretability in terms of coherent structures and involved dynamics. Nevertheless, despite these promising results, the predictive capabilities of Galerkin models were shown to be limited and often lead to inaccurate reduced models. Indeed, while the POD is optimal for the reconstruction error in a suitable norm, the Galerkin projection might not be (\cite{rowley2016review}).

Because of these limitations, some research efforts have been focused on deriving purely data-driven models via sparse regression methods (\cite{brunton2016sindy,loiseau2018constrained}), cluster modeling (\cite{kaiser2014crom}) or Koopman theory (\cite{mezic2013koopman}) to cite just a few. The representation of the dynamics with the Koopman operator has garnered a lot of interest in recent years, both through the DMD algorithm (\cite{Schmid2008dmd}) and its extensions (\eg, EDMD and KDMD \cite{Williams2015edmd,Williams2015kdmd}) which leverage nonlinear state transformations to simplify the modeling problem. 

These purely data-driven works are part of a wider trend towards hybridisation of simulation methods with machine learning techniques. Specifically, neural networks  which have been applied to numerous physical modeling problems, from black box resolutions through graphs networks allowing the resolution of problems on unstructured meshes (\cite{pfaff2021graph}), to more interpretable approaches like the \textit{physics informed neural networks} proposed in \cite{pinns}. Over the last few years, several works have proposed unsupervised learning techniques to avoid the mesh construction (mesh-free methods).  These algorithms train deep neural networks to approximate PDE solutions by satisfying the differential operator, initial conditions and boundary conditions for a specific PDE. \cite{Berg_2018} used fully connected layers to approximate the solution on complex geometries. \cite{raissi2018deep,sirignano2018dgm} discussed the possibility to solve high dimensional problems through neural networks. The authors of \cite{sirignano2018dgm} proved that the neural network converges to the PDE as the number of hidden units increases. A more detailed review on machine learning/CFD hybridization can be found in \cite{brunton2020review}. At the same time, the deep learning community has developed powerful tools by adapting physical modeling concepts to deep learning, for instance the Neural ODE approach was proposed for the modeling of continuous transformations (\cite{chen2019node}), stable neural architectures were also developed by embedding neural networks with invariant structures (\cite{haber2017stable,greydanus2019hnn,behrmann2019invertible}). Neural networks have also been applied in the context of model order reduction, notably to learn useful representations of a system's state (\cite{bollt2017edmd,otto2019lran,duraisamy2020edmd,Erichson2019lyapStable}).

As part of this interest in hybridization, we propose a novel architecture for the development of closures for Galerkin reduced models. Despite their limited performance with nonlinear systems, the physical information embedded in their structure can be retained and used as a basis for the construction of a hybrid reduction approach. Applications of this idea have already been proposed in the literature (\cite{wang2020recurrent}) and are discussed in the following paragraph. We refer the reader to \cite{ahmed2021closures} for a thorough review on the topic. 

\subsection{Related Work}
\label{sec:RelatedWork}
Reduced order modeling methods offer powerful and straightforward industrial applications, thus, their improvement has been the subject of a large body of literature. The proposed solutions can be divided into \textit{intrusive} and \textit{non intrusive} methods. \textit{Intrusive} approaches aim at learning a closure model for the available ROM such that high fidelity data are fitted, while \textit{non intrusive} approaches propose to completely replace pre-existing models with learnt black-box forecasting methods. Several works have proposed to leverage deep learning methods to develop non intrusive forecasting models on low dimensional spaces and represent physical simulation problems (\citep{vlachas2021led,maulik2019latent,pawar2019enabler,fresca2021PODDL,gao2020nonintrusive}). Specifically, the necessity of exploiting temporal information to reconstruct accurate dynamics has been underlined in \cite{bukka2020podlstm,maulik2020advection,eivazi2020unsteady,Wu2019tcn}. More theoretically grounded works have also proposed to develop intrusive non-Markovian closure models for existing Reduced Order Modeling approaches (\cite{wang2020recurrent,Pawar2019recovery}). They are motivated by the Mori-Zwanzig formalism (\citep{zwanzig2001mz}), which provides a theoretical framework for the modeling of partially observed systems. Similar to these efforts, our work aims at developing a theoretically grounded reduced modeling method for the forecasting of physical systems, preserving the physical insights provided by the projected PDE describing the problem. Contrary to most previous works, which are based on the use of recurrent neural networks, our proposal is constructed around a time-continuous memory formulation with numerous advantages over discrete time models. Moreover, we underline below the  higher degree of interpretability of our solution. The main novelties of this work are listed below and Table~\ref{tab:ComparePapers} also provides a summary of the potential crossovers between existing methods and the present work.

\begin{table}[]
    \centering
    \begin{tabular}{| l | c | c | c | c |}
        \hline &Non-Markovian&Non Intrusive&Continuous& Memory \\
         &  &  &  &interpretability\\
        \hline Wang \textit{et al.} (\citep{wang2020recurrent}) & \checkmark &  & & \checkmark\\
        Pawar \textit{et al.} (\citep{Pawar2019recovery}) & \checkmark &  & & \\
        Maulik \textit{et al.}  (\citep{maulik2019latent}) & \checkmark & \checkmark & \checkmark  &\\
        Vlachas \textit{et al.} (\citep{vlachas2018lstm}) & \checkmark & \checkmark & & \checkmark\\
        Wu \textit{et al.} (\citep{Wu2019tcn}) & \checkmark & \checkmark & & \checkmark\\
        Maulik \textit{et al.} (\citep{maulik2020advection}) & \checkmark & \checkmark & &\\
        Pawar \textit{et al.} (\citep{pawar2019enabler}) & \checkmark & \checkmark & & \\
        Our work & \checkmark &  & \checkmark & \checkmark \\ 
        \hline
    \end{tabular}
    \caption{Comparison of closure approaches from the recent literature.}
    \label{tab:ComparePapers}
\end{table}

\begin{itemize}
    \item \textbf{Intrusivity}: Closure modeling has been a topic of interest since the early days of numerical simulation, thus, developing \textit{intrusive} correction models which combine with existing physical models is not a novel approach in and of itself. However, most deep learning approaches do not take this route and propose to learn forecasting models from scratch, ignoring the underlying physical laws. In this work, we show that the Galerkin ROM can be used inside the training loop to optimise the closure model in an \textit{a-posteriori} fashion, that is, by simulating the whole model and assessing its performance. Embedding the existing ROM in the training strategy allows us to leverage pre-existing physical information rather than replace it with a physics-agnostic model.
    
    \item \textbf{Continuity}: The proposed model is embedded with a novel time continuous memory formulation. This increases the applicability of the model as it can be plugged in any initial value problem solver without being biased against specific time-step and/or numerical scheme choices made during training. Moreover, this flexibility implies that this work can be used to model arbitrarily stiff problems through the use of adaptive time-stepping schemes. Finally, the continuous structure allows the model to be used in combination with irregularly spaced data, often encountered in real-life problems, with no additional considerations to interpolation between samples.
    
    \item \textbf{Memory Interpretability}: Contrary to classical recurrent neural networks such as the Long Short Term Memory (LSTM) used in the state of the art, the proposed memory formulation was specifically designed for numerical simulation purposes. Particularly, the time evolution of the memory has a closed form solution, which means that the memory term can be initialised to any desired degree of precision. Our proposed formulation also allows for the evaluation of the \textit{time persistence} of information in memory, increasing the overall interpretability of the model.
    
    \item \textbf{End to End Training}: In contrast with other works which propose to learn the closure model in an \textit{a-priori} fashion, \textit{i.e.} by learning the dynamics correction as a standalone regression problem, we integrate the imperfect model directly within the training strategy. Thus the correction model accurately learns to compensate for the sensitivity of the Galerkin ROM and account for the long term effects of unstable, low energy modes in the system. Indeed, it has been observed in the literature that this strategy lead to more stable and accurate models (\cite{strofer2021turbulence,otto2019lran,kiwon2020solverintheloop}).

\end{itemize}

The outline of the paper is as follows: the POD-Galerkin reduction method is introduced in Section \ref{sec:ModelReduction} and its limitations are discussed. Section \ref{sec:DataDrivenResidual} details our main contribution on the derivation of a continuous in time correction architecture for reduced order models. Motivation for our work through comparisons with existing approaches is provided in Section \ref{sec:Interpretation}, while the selected test cases and results are respectively discussed in Section \ref{sec:CasePres} and \ref{sec:Results}. Section \ref{sec:Conclusion} concludes the paper.

\section{Model reduction approach}\label{sec:ModelReduction}

In this Section we aim at describing the POD-Galerkin model reduction strategy, here illustrated in the context of fluid dynamics. In particular we focus on the incompressible Navier-Stokes equations owing to their relevance in engineering applications and since they represent a challenging test-bed for model reduction techniques. The dynamics of the velocity field $\mathbf{u}(x,t)$ and pressure field $p(x,t)$ are governed by:
\begin{equation}
\label{eq:NS}
    \begin{array}{rcl} 
        \cfrac{\partial \mathbf{u}}{\partial t} + (\mathbf{u}\cdot \nabla) \mathbf{u} & = & -\nabla p +\cfrac{1}{Re} \nabla^2 \mathbf{u} \\ 
        \nabla \cdot \mathbf{u} & = & 0
    \end{array}
\end{equation}
with $Re$ the Reynolds number. Computationally demanding operations are required to numerically solve Eq. \eqref{eq:NS}. The typical approach of POD Galerkin Reduced Order Models (ROMs) is then employed to identify a low-dimensional approximation of the Navier-Stokes equations based on time invariant spatial features. The method, and its derivation for the Navier-Stokes equations, is presented in Section \ref{PODGalerkin}, and its limitations are discussed in Section \ref{drawbacks}. For a deeper introduction to POD Galerkin reduced order models we refer the interested reader to \cite{holmes2012turbulence, lassila2014model}.

\subsection{POD Galerkin models}\label{PODGalerkin}

After gathering data from experiments or the numerical simulation of Eq.~\eqref{eq:NS}, one can use the Proper Orthogonal Decomposition, also known as Principal Component Analysis (PCA), to identify dominant structures in the data. The POD method relies on a matrix, referred to as the snapshot matrix, $\mathbf{U} = \{\mathbf{u}(t_i) \vert i=1, \ldots, n_t\}$ whose columns $\mathbf{u}(t) \in \mathbb{R}^{d n_x}$ correspond to the $d$-dimensional state of the system (for example, the three components of the fluid velocity) indexed upon the $n_x$ points of a spatial grid at time $t$. In the sequel, it is supposed that $n_x \ge n_t$, which is common in numerical simulation applications, but not a requirement of the POD method.

Let us consider the thin singular value decomposition (SVD) of this real-valued matrix: $\mathbf{U = V \Sigma W^\transpose}$, where the matrix $\mathbf{V} \in \mathbb{R}^{d n_x \times n_t}$ and $\mathbf{W}\in \mathbb{R}^{n_t \times n_t}$ respectively hold the left and right singular vectors of $\mathbf{U}$. The diagonal matrix $\mathbf{\Sigma}\in \mathbb{R}^{n_t \times n_t}$ holds the singular values of $\mathbf{U}$, usually arranged such that $\sigma_1 \geq \sigma_2 \geq \ldots \geq \sigma_{n_t} \geq 0$.

Columns of $\mathbf{V}$ are the time-invariant spatial modes defining a POD basis. These POD modes are orthonormal $\langle \mathbf{v}_i , \mathbf{v}_j\rangle = \delta_{ij}$, with $\delta_{ij}$ the Kronecker delta and $\langle\cdot,\cdot\rangle$ is here Euclidean: $\langle \mathbf{v}_i , \mathbf{v}_j\rangle = \mathbf{v}_i^\transpose \, \mathbf{v}_j$. These modes are useful for dimensionality reduction because, for any $r < n_t$, the subspace spanned by the basis $\mathbf{V}_r = \{\mathbf{v}_1, \mathbf{v}_2, \ldots, \mathbf{v}_r\}$ optimally approximates the data $\mathbf{U}$ over the set of $d n_x \times r$ matrices in the sense that it minimizes the reconstruction error $E_r$ defined as:

\begin{equation}
E_r = \Vert \mathbf{U - V}_r \mathbf{V}_r^\transpose \mathbf{U} \Vert_F, \qquad \forall \: r \in \left[1, \ldots, n_t\right].
\end{equation}

This error can be related to the sum of the discarded singular values: $E_r^2 = \sum_{k=r+1}^{n_t}  \sigma_k^2$. This implies that the information captured by the first $r$ modes in the POD basis can be quantified by looking at the following ratio:

\begin{equation}
\label{eq:treshold}
R(r) = \frac{\sum_{k=1}^{r}  \sigma_k^2}{\sum_{k=1}^{n_t}  \sigma_k^2}.
\end{equation}

Introducing a projector $\mathbf{P}= \mathbf{V}_r\mathbf{V}_r^\transpose$ of $\mathbb{R}^{d n_x}$ onto the invariant subspace spanned by the basis $\mathbf{V}_r$, and $\mathbf{Q} = (\mathbf{I}-\mathbf{P})$ its orthogonal complement, the full state vector $\mathbf{u}(t)$ can be decomposed as:

\begin{equation}
\mathbf{u}(t) = \mathbf{Pu}(t) + \mathbf{Qu}(t).
\end{equation}

The key idea is to select a suitable $r$ in Eq.~\eqref{eq:treshold} based on an energy criterion such that $\mathbf{Qu}(t)$ can be neglected for the problem under consideration. The full state $\mathbf{u}(t)$ can then be approximated as: 
\begin{equation}
\label{eq:ReducedForm}
\mathbf{u}(t) \approx \mathbf{\widetilde{u}}(t) = \mathbf{V}_r \mathbf{a}(t)
\end{equation}
where $\mathbf{a}(t):= \mathbf{V}_r^\transpose \mathbf{u}(t)$ is the vector containing temporal coefficients associated with the spatial modes in $\mathbf{V}_r$. From the approximation defined in Eq.~\eqref{eq:ReducedForm}, model reduction arises whenever $r \ll n_x$. Under this modal decomposition \textit{ansatz}, the original spatio-temporal problem of solving the PDE is reduced to the computation of the temporal coefficients $\mathbf{a}(t)$ as the spatial information is expressed in the invariant POD basis $\mathbf{V}_r$. However, to achieve a gain in computational cost, one needs a way to compute the reduced coordinates vector without accessing the full order solution. This is where the Galerkin projection method can be used to obtain an equation for the reduced coordinates. The method is based on the reduction of the original model equation:

\begin{equation}\label{FullPDE}
\dv{\mathbf{u}(t)}{t} = \boldsymbol{f}(\mathbf{u}(t))    
\end{equation}
where $\boldsymbol{f}$ corresponds to the full order dynamics discretised in space. By projecting this equation onto the time-independent POD basis, a dynamical model for the reduced coordinates is obtained:

\begin{align}\label{eq:PGROM}
    &\dv{\mathbf{a}(t)}{t} = \mathbf{V}_r^\transpose \boldsymbol{f}(\mathbf{u}(t)).
\end{align}

With this Galerkin projection, the dimension of the problem is reduced but the model still requires knowledge of the full state $\mathbf{u}(t)$ to be resolved. To obtain a model that solely depends on the coordinates $\mbf{a}(t)$, the dynamics $\boldsymbol{f}$ of the model have to be evaluated from the approximate form of the state (see Eq.~\eqref{eq:ReducedForm}). This leads to the following system of $r$ ODEs:

\begin{equation}\label{eq:ApproxProjection}
    \dv{\mathbf{a}(t)}{t} \approx \mathbf{V}_r^\transpose \boldsymbol{f}(\mathbf{V}_r\mathbf{a}(t)).
\end{equation}

If $\boldsymbol{f}$ is the right hand side of the incompressible Navier-Stokes in Eq. (\ref{eq:NS}) and the POD basis is retrieved from the velocity field snapshot matrix, the approximation in (\ref{eq:ApproxProjection}) leads to the following ROM:

\begin{equation}\label{eq:TensorialROM}
    \dv{a_i(t)}{t} = \sum_{j=1}^r \mathcal{\widetilde{L}}_{i,j} \, a_i(t) + \sum_{j=1}^r \sum_{k=1}^r \mathcal{\widetilde{Q}}_{i,j,k} \, a_j(t) \, a_k(t), \qquad \forall \, i = 1, \ldots, r
\end{equation}
where tensors $\widetilde{\mathcal{L}}$ and $\widetilde{\mathcal{Q}}$ are reduced versions of the dissipative and advective parts of the original Navier Stokes equations. Because operators $\mathcal{L} = \Delta \mbf{u}(x,t)$ and $\mathcal{Q}=\nabla(\mbf{u}(x,t)\mbf{u}(x,t))$ are spatial operators (here $\mbf{u}(x,t) \in \mathbb{R}^d$ refers to the velocity vector), they can be pre-assembled from the POD modes. In contrast with this favorable situation, other PDE problems might necessitate that the dynamics be evaluated from the reconstructed solution $\mbf{u}(t) \approx \mathbf{V}_r\mathbf{a}(t)$, which can imply a significant cost and require additional reduction work, \cite{chaturanbaut2009deim}. It is worth noting that the pressure term is not present in Eq.~\eqref{eq:TensorialROM}. This is due to the fact that, in the full order model, the pressure term accounts for the incompressibility condition ($\nabla \cdot \mbf{u} = 0$) and can be interpreted as a Lagrange multiplier used to ensure this constraint during the simulation.

Because the reconstructed solution is a linear combination of the POD modes which are all divergence-free, the incompressibility condition is always satisfied and the pressure term can then be ignored without significant accuracy loss (\cite{rowley2016review,Noack2003cylinder}).

\subsection{Limitations of POD-Galerkin models}\label{drawbacks}

Because POD-Galerkin models rely on the projection of the original model equations, they retain part of the structure from the full order physical model. This makes them relatively robust to new initial conditions comparatively to purely data-based approaches which can struggle to model conditions unseen during training, as well as suffer from overfitting depending on the way \EMmodif{}{they were trained}. It should be noted however that approximation errors are embedded by construction in their structure. Equation (\ref{eq:PGROM}) shows that an exact expression for the dynamics of the reduced coordinates is obtained by projection of the model equation on the POD basis. If the orthogonal complement to $\mathbf{V}_r$ is not neglected, equation (\ref{eq:PGROM}) reads: 
\begin{equation}\label{eq:ResidualDerivation}
    \dv{\mathbf{a}(t)}{t} =  \mathbf{V}_r^\transpose \boldsymbol{f}(\mathbf{u}(t))   =  \mathbf{V}_r^\transpose \boldsymbol{f}(\mathbf{Pu}(t) + \mathbf{Qu}(t)),
\end{equation}
with $\Vert \mathbf{Qu}(t) \Vert_F \ll \Vert \mathbf{Pu}(t) \Vert_F$ if $r$ is large enough. Considering a first order approximation of Eq. (\ref{eq:ResidualDerivation}), the temporal dynamics of the POD coefficients can be evaluated from:

\begin{equation}
    \label{eq:}
    \dv{\mathbf{a}(t)}{t}  \approx \mathbf{V}_r^\transpose \boldsymbol{f}(\mathbf{V_r \, a}(t)) + \underbrace{\mbf{V}_r^\transpose \mbf{J}_{\mbf{Pu}(t)} \mbf{Qu}(t)}_{\mathcal{R}} 
\end{equation}
with $\mbf{J}_{\mbf{Pu}(t)}$ being the Jacobian of the discretised nonlinear system $\boldsymbol{f}$, evaluated at the state $\mbf{Pu}(t)$. The residual term $\mathcal{R}$ accounts for the \textit{dynamics of the complement of the POD basis} in the space spanned by the POD basis. Classical POD-Galerkin reduced models ignore this error term and directly compute the dynamics of the model from the approximate solution as presented in Section~\ref{PODGalerkin}.

This approximation means that small errors on the dynamics will compound over time and lead to significant discrepancies between the true and simulated trajectories. This is especially true in the case of nonlinear dynamical systems where orthogonal projection on the POD basis can suppress an important part of the dynamics. In the case of the Navier Stokes equations, POD-Galerkin models have been shown to fail to reproduce the dynamics even in simple cases like the flow over a cylinder \cite{Noack2003cylinder}. It should be noted that different projection schemes have been proposed to alleviate this issue (\cite{carlberg2018petrov}), however, these often require additional considerations on the construction of the reduced model. 

The aim of this paper is to retain the simplicity of the POD-Galerkin method, and learn the \textbf{complement} of the \textbf{ROM} using \textbf{deep} learning methods, thus, we call our proposed method \textbf{Complemented Deep - Reduced Order Model}.

\subsection{Non-Markovianity and Takens' theorem}

\begin{figure}
    \centering
    \includegraphics[width=0.95\textwidth]{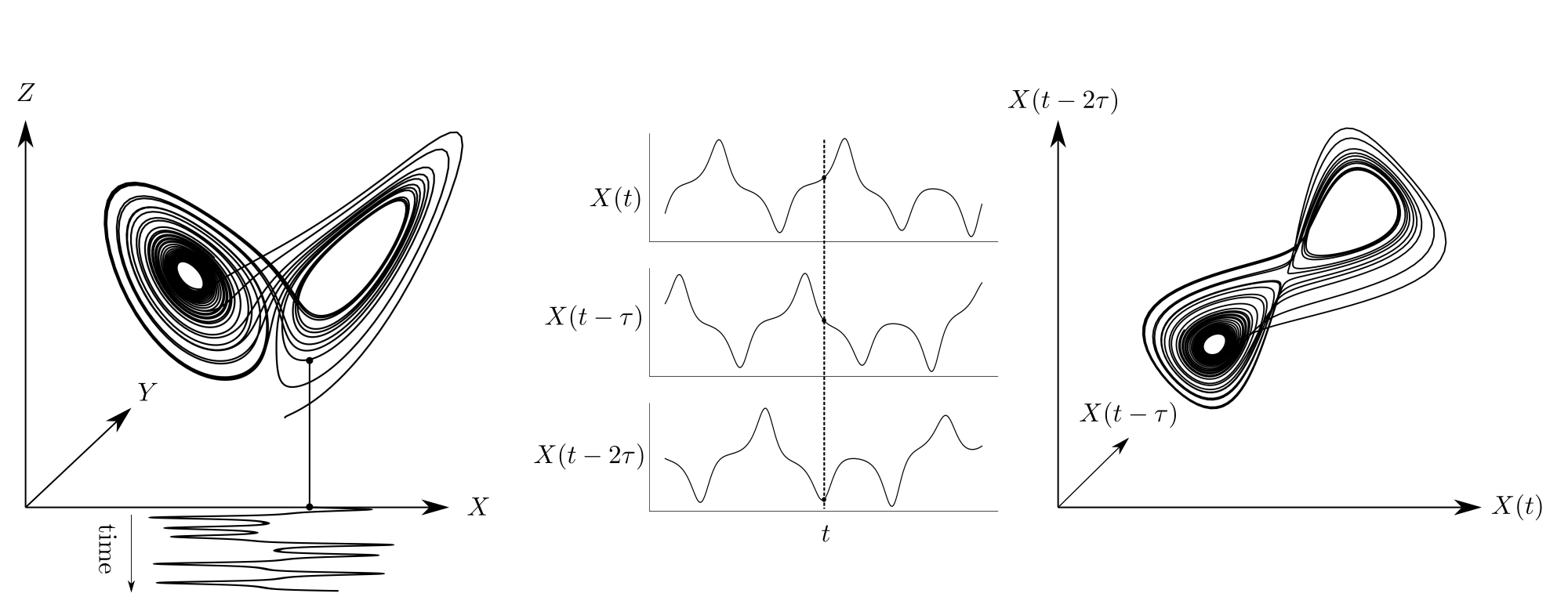}
    \caption{Illustration of the Takens' delay embedding theorem. Left: Original Lorenz attractor; middle: Delayed time series of the $X$ coordinate; right: Reconstructed attractor.}
    \label{fig:LorenzTakens}
\end{figure}

The residual depends on information from a subspace orthogonal to the span of the POD basis. This means that an accurate correction model cannot be directly computed from the reduced state $\mbf{a}(t)$. However, we leverage the fact that the information lost by projection of the full order state can be retrieved by considering past states of the system. This hypothesis is formalized by the Takens' theorem (\cite{takens1981takens}), which states that, under mild conditions, the dynamics of a state vector can be reproduced by constructing a time-embedding from time-lagged observables: $\mbf{z}(t) = \Big(z(t), z(t-\tau), z(t-2\tau),\ldots,z(t-k\tau)\Big)$, with $k$ large enough.\footnote{We tacitly assumed here that $z$ is a suitable observable. Observability analysis goes beyond the scope of this paper. Nevertheless, it is well known that higher harmonic POD modes are enslaved to dominant POD modes \cite{loiseau2018constrained,callaham2021role}. Therefore, in the rest of the paper, we will assume that the dynamics of the unobserved space spanned by the columns of $\mathbf{Q}$ can be retrieved from past observations of the dynamics in the space defined by $\mathbf{P}$.} This is illustrated in Figure \ref{fig:LorenzTakens}, which shows the Lorenz attractor observed via its embedded $X$-component. By constructing a 3-dimensional embedding of the obtained time-series, an attractor is obtained, which preserves the topology of the true attractor (\eg, symmetries, correlation dimension, etc.).

This suggests that the correction term $\mathcal{R}$ in POD-Galerkin reduced order models is non-Markovian and should consider past states of the system $\Big(\mbf{a}(t-\tau), \mbf{a}(t-2\tau), \ldots\Big)$. However, such a discrete time-embedding is not well suited in the context of continuous time models such as equation \eqref{eq:TensorialROM}. Indeed, using discrete time steps in combination with adaptive step time marching schemes, such as the Runge-Kutta method \cite{RK45}, would require specific considerations about interpolation between the simulation steps and the required embedding steps. To address this time-continuity issue while retaining a non-Markovian correction structure, we propose to use delay differential equations (DDE) with a continuous embedding of the past information:

\begin{equation}\label{eq:DDE}
    \dv{}{t}\mbf{a}(t) = \boldsymbol{g}(\mbf{a}(t),\mbf{y}(t)) \quad , \quad \mbf{y}(t)=\int_{-\infty}^t e^{(\tau-t)\lambda}\mbf{a}(\tau)\ddroit\tau
\end{equation}
with $\lambda \in \mathbb{R}_+$ sufficiently large for the integral above to be bounded.

These equations retain information from past states of the system in a time-continuous manner and are used for a number of modeling applications such as epidemiology or population dynamics (\cite{Gopalsamy1992popdyn,salpeter1998tuberculosis}). In this formulation, the dynamics $\boldsymbol{g}$ depend both on the partially observed states of the system $\mbf{a}(t)$ and a memory variable $\mbf{y}(t)$, corresponding to the integral of past observables, damped in time by an exponential decay. In fact, the memory term $\mbf{y}$ can be defined in many ways, depending on the problem at hand. However, the exponential decay formulation was chosen because \textit{i}) it provides the model with the ability to consider recent states of the system while older observations are discarded and \textit{ii}) it can be solved by directly augmenting the original system with a second ODE for the memory:

\begin{equation}\label{eq:DDEODE}
\begin{array}{ccl}
    \displaystyle \dv{}{t}\mbf{a}(t) &=& \boldsymbol{g}(\mbf{a}(t),\mbf{y}(t)),\\
    \displaystyle \dv{}{t}\mbf{y}(t) &=& \mbf{a}(t) - \lambda \mbf{y}(t).
\end{array}
\end{equation}

The exponential decay $e^{(\tau - t)\lambda}$ acts as a filter of width $1/\lambda$ on the observables evolution and we show in a later section that the value of the decay rate ($\lambda$) can be learned in a data driven setting. Other applications of the same augmented ODE exist in literature but with different purposes such as modelling of subgrid-scales in Large-Eddy Simulations (LES) \cite{pruett2003temporally} or to find unstable-steady solutions of the Navier-Stokes equations \cite{aakervik2006steady}. \EMmodif{}{As mentionned earlier, we have made the convenient choice of the exponential kernel since it can be described by a linear ODE and easily constrained to model dissipative dynamics, but alternative kernels exist in the DDE literature which could be considered if the exponential kernel became too constraining for certain cases.}

It is worth noting that the memory variables $\mbf{y}$ have the same dimension as the observations $\mbf{a}$ in (\ref{eq:DDEODE}). This limitation might introduce a significant information bottleneck in the model. Indeed, Takens' theorem states that the dimension required to obtain a satisfactory embedding can go as high as twice the intrinsic dimension of the true attractor. Although there is no similar result for the continuous case, the limited dimension of the memory $\mbf{y}$ may prevent deriving an accurate correction model. As a result, we define an encoding map $\mbf{E}: \mathbb{R}^r \to \mathbb{R}^{n_E}$, used to lift the observations $\mbf{a}$ to a higher dimensional space to increase the dimension of the memory:

\begin{equation}
\mbf{y}(t) = \int_{-\infty}^t e^{(\tau-t)\lambda} \mbf{E}(\mbf{a}(\tau))\ddroit\tau.
\label{eq:AugDDE}
\end{equation} 

\EMmodif{}{In fact, the use of such an encoding map to unfold non-linear dynamics and recover a linear ODE is rooted in Koopman theory as each encoded coordinate can be considered as an observable of the original state. While approaches such as dictionnary learning (\cite{Williams2015edmd}) and kernel methods (\cite{Williams2015kdmd}) have been proposed to learn these observables, we use neural networks to avoid additional optimization considerations and retain flexibility, similar to the works of \cite{bollt2017edmd,duraisamy2020edmd}.} Using the modified DDE architecture \EMmodif{}{(Eq.\eqref{eq:AugDDE})} to close reduced order models, the correction operator $\mathcal{R}$ acting on the memory $\mathbf{y}(t)$ becomes an application from memory space to phase space: $\mathbb{R}^{n_E} \to \mathbb{R}^r$. Finally, the following augmented reduced order model is obtained:

\begin{equation}\label{EncDDEODE}
\begin{array}{c l c c c}
    \displaystyle \dv{}{t}\mbf{a}(t) &=& \mbf{V}_r^\transpose \boldsymbol{f}(\mbf{V}_r \mbf{a})& +& \mathcal{R}(\mbf{y}), \\
    \displaystyle \dv{}{t}\mbf{\mbf{y}}(t) &=& \mbf{E}(\mbf{a}) &-& \lambda \mbf{y}.
\end{array}
\end{equation}

\begin{figure}
    \centering
    \includegraphics[width=0.95\textwidth]{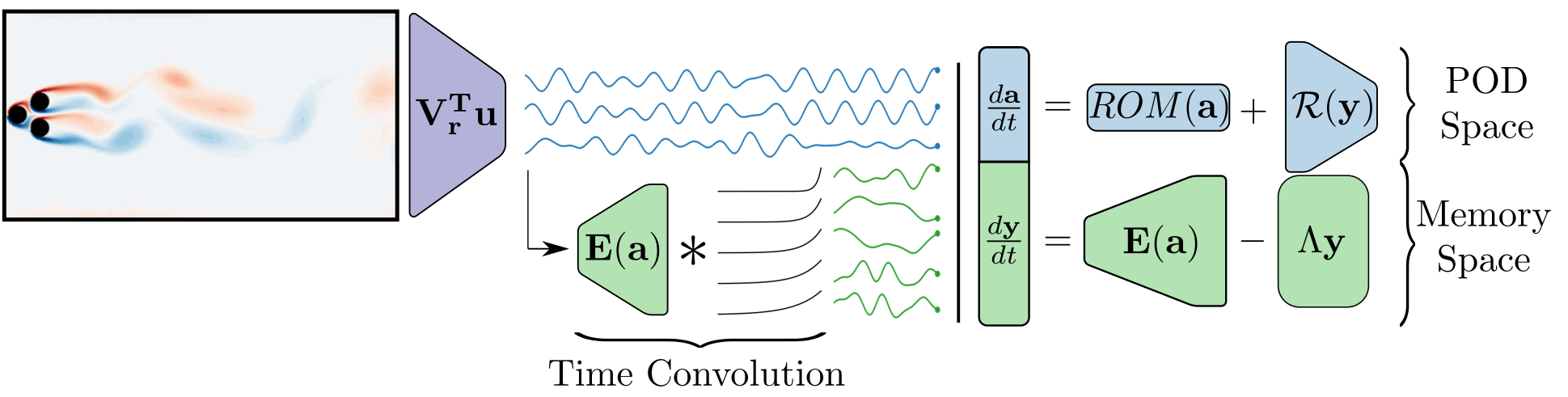}
    \caption{Visualisation of the CD-ROM approach. The full order solution snapshots are projected on the POD basis to obtain time series of the reduced coordinates ($\mbf{a}$). These reduced coordinates are then augmented with a memory $\mbf{y}(t)$. Finally, the augmented dynamical model is presented on the right hand side of the image.}
    \label{fig:RecapCDROM}
\end{figure}

\EMmodif{}{This proposed augmented ROM architecture has a similar form to the Mori-Zwanzig formalism \cite{zwanzig2001mz} which derives an equation for the dynamics of partially observed systems. Although the parallel with the CD-ROM approach is not formally established, we discuss these similarities in section \ref{sec:MZ} as it provides additional motivation for the above choices. Finally, the augmented ROM formulation is }summarised in Figure \ref{fig:RecapCDROM} to help illustrate the idea.

\section{Data driven learning of the residual}\label{sec:DataDrivenResidual}

Although the motivation for the existence of the residual $\mathcal{R}$ and encoding map $\mbf{E}$ was outlined above, little information about their form can be derived from the previous expressions. In this context, we leverage the universal function approximator property of artificial neural networks \cite{leshno1993approximator} to model the missing terms.

\subsection{Neural Networks}\label{NeuralNetworks}

Our work relies on feed-forward neural networks with a very simple structure. These networks can be used to learn any smooth nonlinear continuous application $\mbf{\Psi}: \mathbb{R}^r \to \mathbb{R}^k$ by optimising the weights of a sequence of $L$ layers:

\begin{align}
    \mbf{\Psi}(\mbf{a}) &= \boldsymbol{\eta}_L \circ \boldsymbol{\eta}_{L-1} \circ \cdots \circ \: \boldsymbol{\eta}_1(\mbf{a})\\
    \boldsymbol{\eta}_l &= \sigma_l(\boldsymbol{W}_l \boldsymbol{\eta}_{l-1} + \boldsymbol{b}_l), \qquad \forall \: l \in \left\{2, \ldots, L\right\}
\end{align}
where $\sigma_l$ is a nonlinear activation function, and the dimension of $\boldsymbol{\eta}_l$ corresponds to the number of neurons in the layer $l$. It has been shown that, provided the dimension of the layer is high enough, the trainable parameters $\boldsymbol{W}_l$ and $\boldsymbol{b}_l$ can be optimised to approximate any function, \cite{leshno1993approximator}. The encoder $\mbf{E}(\mbf{a})$ and the residual $\mathcal{R}(\mbf{y})$ are both approximated with neural networks with parameters $\boldsymbol{\theta}_\mbf{E} = \Big(\{\boldsymbol{W}_l,\boldsymbol{b}_l\},\: l=1,..,L_\mbf{E}\Big)$ and $\boldsymbol{\theta}_\mathcal{R}$ respectively.

\subsection{Memory time scales}\label{timescales}

\begin{figure}
    \centering
    \includegraphics[width=0.95\textwidth]{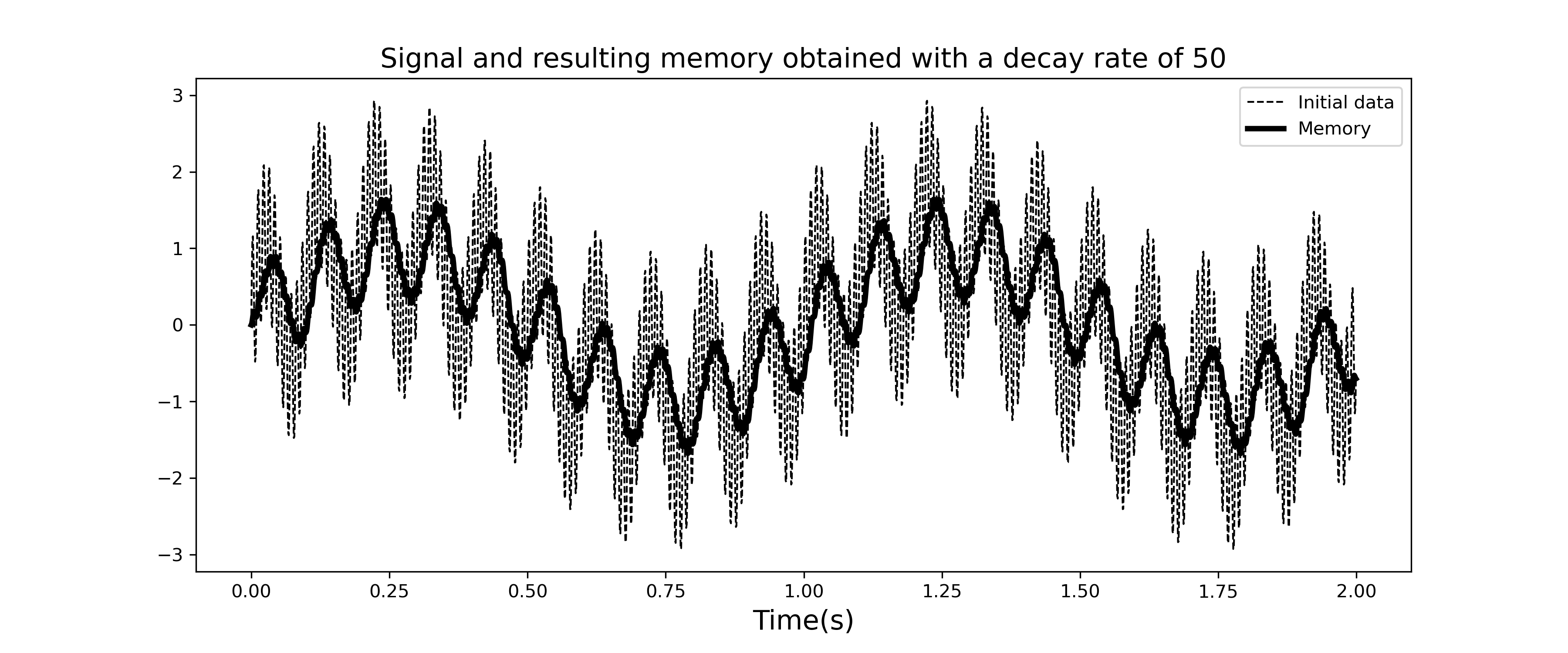}
    \caption{Superposition of sinusoidal signals of frequencies 1, 10 and 100 Hz, filtered by an exponential decay with a decay rate $\lambda=50$. The higher frequency (100 Hz) is filtered out while the other two are reproduced in the memory signal. This implies that recent events in memory due to the $100$ Hz frequency are filtered out so that the memory is mostly driven by  \textit{older} events associated with lower frequencies. Note that the memory signal has been scaled by a factor $\lambda$ for clarity.}
    \label{fig:lowPass}
\end{figure}

Physical systems often involve a variety of phenomena each evolving at different time scales. Capturing these phenomena can be critical to accurately model these systems, which is why the memory should be able to retain information at different rates. The time scales accounted for by the memory are driven by the parameter $\lambda$ of the exponential decay in Eq.~\eqref{eq:DDE}, which acts as a low-pass filter on the encoded trajectory (see Figure \ref{fig:lowPass} for an illustration). 

To retain information at different rates, $\lambda$ can be adjusted for each observable $E_i(\mbf{a})$.
Equation \eqref{EncDDEODE} is then modified accordingly with the single $\lambda$ parameter replaced with a diagonal matrix $\mbf{\Lambda}\in \mathbb{R}^{n_E}_+$ whose entries can be learned to select time scales relevant to the system at hand. This finally results in the CD-ROM architecture:

\begin{equation}\label{eq:DeepROM}
\begin{array}{c l c c c}
    \dv{}{t}\mbf{a}(t) &=& \mbf{V}_r^\transpose \boldsymbol{f}(\mbf{V_r a})& +& \mathcal{R}(\mbf{y};\boldsymbol{\theta}_\mathcal{R}), \\
    \dv{}{t}\mbf{\mbf{y}}(t) &=& \mbf{E(\mbf{a};\boldsymbol{\theta}_E}) &-& \mbf{\Lambda} \, \mbf{y}.
\end{array}
\end{equation}

\subsection{Training strategy}\label{NODE}

Optimising the parameters of the model \eqref{eq:DeepROM} requires consideration of past states of the system as well as their impact on the dynamics in the future. In some sense, the problem is similar to the optimisation of classical recurrent neural networks for the simulation of dynamical systems \cite{vlachas2018lstm}. The major difference is that the aim of the present work is to derive a continuous time dynamical model, while recurrent networks have traditionally been used to model transitions between discrete time instants. As a result, we cannot use standard backpropagation through time to optimise the model. Instead, the NeuralODE approach \cite{chen2019node} is used. Initially proposed as the continuous equivalent of residual networks (\cite{he2015resnet}), 
Neural ODEs are based on the well established adjoint backpropagation method (\cite{pontriagin1962adjoint}). It is briefly presented here while a more detailed derivation is given in \ref{AdjointAnnex}.
The adjoint backpropagation method can be used to solve constrained optimization problems of the form:

\begin{align}
    \min_{\mbf{\theta}} \quad & J(\mbf{x}(T;\mbf{\theta}))\\
    \textrm{s.t.} \quad & \dv{}{t}\mbf{x}(t) = \boldsymbol{f}(\mbf{x};\mbf{\theta})\\
    & \mbf{x}(0) = \mbf{x}_0.
\end{align}

The method allows for the computation of the gradient $\dv{}{\mbf{\theta}}J$, at roughly twice the cost of integrating forward in time from $t=0$ to $t=T$, through the following expression:

\begin{equation}
    \dv{}{\mbf{\theta}}J = \int_0^T -\mbf{\mu}^\transpose \frac{\partial \boldsymbol{f}}{\partial \mbf{\theta}} \ddroit t
\end{equation}
where the adjoint state $\mbf{\mu}$ can be computed by integrating the adjoint ODE:

\begin{align}
    \dv{}{t}\mbf{\mu^\transpose}(t) &= - \mbf{\mu^\transpose}\frac{\partial \boldsymbol{f}}{\partial \mbf{x}}\Big|_t, \label{eq:AdjointODE}\\
    \mbf{\mu}^\transpose(T) &= \frac{\partial J}{\partial \mbf{x}} \Big|_T.
\end{align}

Computing the gradient of $J$ with respect to the parameters thus requires integration of the adjoint ODE backwards in time, from $t=T$ to $t=0$. If the dynamics $\boldsymbol{f}$ are expressed by a Neural Network, the required vector-Jacobian products $\mbf{\mu}^\transpose\frac{\partial \boldsymbol{f}}{\partial \mbf{x}}\Big|_t$ and $\mbf{\mu}^\transpose \frac{\partial \boldsymbol{f}}{\partial \boldsymbol{\theta}}\Big|_t$ can seamlessly be evaluated through automatic differentiation by any classical deep learning framework.

As explained in Section \ref{PODGalerkin}, reduced models of the incompressible Navier-Stokes equations can be expressed directly in terms of the reduced coordinates $\mbf{a}$ and evaluated through simple tensorial expressions (see equation (\ref{eq:TensorialROM})). This means that the ROM dynamics can be directly computed and back-propagated through once the reduced operators $\widetilde{\mathcal{L}}$ and $\widetilde{\mathcal{Q}}$ are assembled (see equation \ref{eq:TensorialROM}). The Deep ROM architecture (\ref{eq:DeepROM}) can then be optimised within the Neural ODE framework by concatenating the reduced state and the memory into a single state vector $\mbf{z(}t) = [\mbf{a}(t), \mbf{y}(t)]$, with dynamics $\boldsymbol{f}(\mbf{z};\mbf{\boldsymbol{\theta}_E,\boldsymbol{\theta}_\mathcal{R},\mbf{\Lambda}})$:

\begin{equation}\label{eq:ArchODE}
\boldsymbol{f}(\mbf{z};\mbf{\boldsymbol{\theta}_E,\boldsymbol{\theta}_\mathcal{R},\mbf{\Lambda}}) = \left(
\begin{array}{l c l}
    \widetilde{\mathcal{L}}\mbf{a} + \mbf{a}^\transpose\widetilde{\mathcal{Q}}\mbf{a} & + & \mathcal{R}(\mbf{y;\boldsymbol{\theta}_\mathcal{R}})  \\
    \mbf{E(a;\boldsymbol{\theta}_E}) & - & \mbf{\Lambda} \mbf{y}
\end{array}
\right).
\end{equation}
    
Since the last modes in the POD basis $\mathbf{V}$ are typically associated to low-energy dissipative scales \cite{ahmed2021closures}, if $r$ is not large enough, the ROM model retrieved by $\mathbf{V}_r$ might be unstable and diverge after a few time integration steps.
This potentially unstable primal model is however embedded in the optimization framework so that the residual is trained accordingly, resulting in a stable, accurate, model.

\subsection{Training Data and Memory Initialisation}

To optimise the parameters of equation (\ref{eq:ArchODE}) through the Neural ODE approach, one only needs knowledge of the true trajectory of the reduced coordinates $\mbf{a}^\star(t)$. As presented in Section \ref{PODGalerkin}, this trajectory data can be obtained by projecting the solutions computed with the full order solver on the POD basis:

\begin{equation}
\mbf{\mbf{a}^\star}(t) = \mbf{V}_r^\transpose \mbf{u}(t).
\end{equation}

If the snapshots are sampled with a time interval $\Delta_t$, simulating the corrected ROM (\ref{eq:ArchODE}) for $n_t$ time steps leads to the following $L^2$ mean squared error:

\begin{equation}\label{eq:trajectory_loss}
    J = \frac{1}{n_t}\sum_{i=1}^{n_t}  \left\|\mbf{a}(i\,\Delta_t) - \mbf{a}^\star(i\,\Delta_t)\right\|_2^2
\end{equation}
for which a gradient can be computed by integrating the adjoint equation (\ref{eq:AdjointODE}). It should be noted that the memory term needs to be initialised by evaluating the memory integral at time $t=0$ which requires knowledge of the past of the true trajectory:

\begin{equation}
    \mbf{y}(0) = \int_{-\infty}^0 \mbf{E(\mbf{a}^\star(\tau);\theta_E)} e^{\mbf{\Lambda}\tau}\ddroit\tau.
\end{equation}

 To be able to compute this integral, the infinite horizon of the memory integral can be relaxed by defining a finite $\tau_{\mathrm{min}}$ from the longest time scale $\lambda_\mathrm{min}$ of the matrix $\mbf{\Lambda}$ and a threshold $\epsilon \ll 1$, chosen to be small enough such that the relative error made on the initial memory is sufficiently small:

\begin{align}\label{Taumin}
    \tau_{\mathrm{min}} & =  -\frac{\log{\epsilon}}{\lambda_\mathrm{min}},\\
    \mbf{y}(0) &\approx \int_{-\tau_{\mathrm{min}}}^0 \mbf{E(\mbf{a}^\star(\tau);\theta_E)}e^{\mbf{\Lambda}\tau}\ddroit\tau.
    \label{eq:memInit}
\end{align}

Because this initial memory term directly depends on the parameters of the encoder and the matrix $\mbf{\Lambda}$, it needs to be re-computed at each training epoch. This can be done very efficiently on a GPU through a simple trapezoidal approximation of the integral in Eq.~\eqref{eq:memInit}. It should be noted that, while the initialization of the memory is necessary to obtain an exact model of the system at hand, it could limit the applicability of the method to certain real life settings. However, there are ways to make it less critical, such as initializing the memory with white noise during training. This would of course impact the accuracy of the model, depending on the system at hand. 

To conclude this section, algorithm \ref{alg:SimplTraining} summarises the NeuralODE training procedure.

\begin{algorithm}
\caption{Training the CD-ROM as a Neural ODE}\label{alg:SimplTraining}
\begin{algorithmic}
\Require{$f(\mbf{z;\theta})$ the dynamics of the CD-ROM, $J$ the loss, $\eta$ the learning rate}
\For{$i \gets 1$ to $N$} \Comment{Training iterations}

\State{Sample a batch of trajectories $\mbf{a}_{-\tau_{\textit{min}}\xrightarrow{}T}^\star$}

\State{Compute the initial memory $\mbf{y_0}$} \Comment{Equation \ref{eq:memInit}}

\State{$\mbf{z_0} \gets [\mbf{a_0^\star,y_0}]$}

\State{$[\mbf{a}_t,\mbf{y}_t] \gets \mbf{z_0} + \int_0^{t}f(\mbf{z;\theta}) \ddroit t$} \Comment{CD-ROM simulation}

\State{Compute the loss $J(\mbf{a}_{0\xrightarrow{}T},\mbf{a}^\star_{0\xrightarrow{}T})$}

\State{$\mu_T \gets \frac{dJ}{d\mbf{z}_T}$} \Comment{Adjoint Initial Condition}

\State{$\mu_t \gets \mu_T - \int_T^t \mu^\transpose \frac{\partial \mbf{f}}{\partial\mbf{z}_t} \ddroit t$} \Comment{Adjoint Equation}

\State{$\dv{}{\mbf{\theta}}J \gets \int_0^T -\mbf{\mu}_t^\transpose \frac{\partial \boldsymbol{f}}{\partial \mbf{\theta}} \ddroit t$} \Comment{Compute the Gradient}

\State{$\theta \gets \theta - \eta \dv{}{\mbf{\theta}}J$}

\EndFor
\end{algorithmic}
\end{algorithm}

\section{Interpretation of the model}\label{sec:Interpretation}

Before presenting the selected test cases and results, further justification and insights into the model are discussed in this Section. First, the Mori-Zwanzig formalism is introduced to frame our model in the context of dynamical systems theory. Then, we study how it can be compared to purely data driven approaches, such as reservoir computing.

\subsection{The Mori-Zwanzig formalism}
\label{sec:MZ}
The Mori-Zwanzig formalism (\cite{mori1965transport,zwanzig1973langevin,zwanzig2001mz}) provides a closed form for the dynamics of partially observed systems by distinguishing three separate terms:

\begin{equation}\label{eq:MZ}
    \dv{}{t}\mbf{\mathbf{a}}(t) = \Omega(\mbf{a}(t)) + \int_0^t K(\mbf{a}(s)) \ddroit s + F(t)
\end{equation}
where $\mbf{a}(t)$ are observables of a system defined as a projection of the full order state onto the observable space and $\Omega$ is the projected part of the original dynamics. These two quantities can respectively be identified as the reduced coordinates and dynamics of classical POD-Galerkin models. The remaining terms account for the impact of the non-observed coordinates of the system on the resolved dynamics. $K$ represents the dynamical exchanges between resolved and unresolved dynamics during the simulation, while $F$ accounts for the incomplete knowledge of the initial condition \EMmodif{}{and system dynamics. Under the condition that the unresolved dynamics be dissipative, which is reasonable when the unobserved coordinates correspond to the small scales of a dynamical system, and that the boundary of the integral term in Eq.~\eqref{eq:MZ} be modified to $-\infty$, the last term in the Mori-Zwanzig formalism vanishes, leading to the following formulation:}

\begin{equation}\label{eq:MZ2terms}
    \dv{}{t}\mbf{\mathbf{a}}(t) = \Omega(\mbf{a}(t)) + \int_{-\infty}^t K(\mbf{a}(s)) \ddroit s.
\end{equation}

The residual model $\mathcal{R}$ proposed above can then be identified with the memory integral defined by the time-convolution kernel $K$, providing a strong connection with our non-Markovian correction hypothesis. Framing the CD-ROM architecture in the context of the Mori-Zwanzig formalism further justifies our modeling choices. Yet, it does not provide additional insights into the form of the correction model since little information is known about the convolution operator which can be infinite dimensional in certain cases. Choices about the structure of $K$ need to be made. In this work, assumptions are made about the vanishing impact of past states of the system on the residual model, accounted for by the matrix $\mbf{\Lambda}$ in our approach.

\subsection{Deep Learning interpretation of the model}

In this paragraph, the link between the CD-ROM architecture and classical deep learning models is discussed. Numerous methods have been proposed to model sequential data, each relying on a specific mechanism to extract and retain meaningful information from the past states of the system. The most popular architectures, GRU and LSTM (\cite{cho2014gru,hochreiter1997lstm,chung2014gated}), both use a combination of \textit{gating} mechanisms to learn long term dependencies in a sequence. Other approaches like reservoir computing rely on an underlying dynamical system forced by the sequence data to predict the required output. In fact, strong similarities can be identified between our approach and \textit{echo state networks} (ESN \cite{Lukosevicius2009reservoir,magri2020esn}), a widespread reservoir computing architecture. ESNs are based on the simulation of random dynamics described by matrices $\mbf{W}_\mathrm{in}$ and $\mbf{W}_R$. The matrix $\mbf{W}_\mathrm{in}$ is used to encode some data $\mbf{x}_t$ in higher dimension, while the matrix $W_R$ holds the weights of the reservoir used to advance the state (memory) $\mbf{y}_t$ in time:

\begin{equation}\label{ESN}
    \mbf{y}_{t+1} = \sigma(\mbf{W}_\mathrm{in} \mbf{x}_t + \mbf{W}_R \mbf{y}_t).
\end{equation}

For the sake of comparison, the nonlinear activation function $\sigma$ is dropped, and we consider that equation (\ref{ESN}) results from the Euler integration with time-step $\Delta_t$ of a continuous system describing the dynamics of the memory $\dv{\mbf{y}}{t}$:

\begin{align}
     \mbf{y}_{t+1} &= \mbf{W}_\mathrm{in} \mbf{x}_t + \mbf{W}_R \mbf{y}_t\\
     &= \mbf{y}_t + \Delta_t \dv{\mbf{y}}{t}
\end{align}
which leads to the following expression:

\begin{equation}
    \dv{\mbf{y}}{t} = \frac{\mbf{W}_\mathrm{in}}{\Delta_t} \mbf{x}_t + \frac{(\mbf{W}_R - I)}{\Delta_t} \mbf{y}_t.
\end{equation}

Because the spectral radius of the matrix $W_R$ is constrained to be less than unity, all the eigenvalues of the operator $\overline{\mbf{W}}_R = \frac{(\mbf{W}_R - \mbf{I})}{\Delta_t}$ have negative real parts, leading the ESN to have memory dissipation properties similar to those of our model. To underline this similarity, we can express the state of the ESN at time $t$ by diagonalising the operator $\overline{\mbf{W}}_R = \mbf{P}\mbf{\Lambda}_R\mbf{P}^{-1}$:

\begin{equation}
    \mbf{y}(t) = \int_0^T \mbf{P}e^{\mbf{\Lambda}_R (t-\tau)}\mbf{P}^{-1}\frac{\mbf{W}_\mathrm{in}}{\Delta_t}\mbf{x}(\tau) \ddroit\tau.
\end{equation}

Thus, a parallel between our model and the ESN is outlined.  Major differences remain in the fact that the dynamics ($\mbf{W}_R$) and encoding matrix ($\mbf{W}_\mathrm{in}$) are not optimised during the ESN training. It should also be noted that nonlinearity is introduced in the ESN dynamics through the $\sigma$ activation function, while our model is based on linear memory dynamics and on a nonlinear encoding operator. These comparisons between direct deep learning methods and the continuous correction approach help build intuition about the role of each term in the model. The encoder can be compared with the input gate of an LSTM, or the $\mbf{W}_\mathrm{in}$ matrix of the ESN, the $\mbf{\Lambda}$ matrix provides a tunable \textit{forget} mechanism, while the residual term $\mathcal{R}$ plays the role of the output operator in our ``continuous recurrent network''.

\section{Case presentation and reduced models}\label{sec:CasePres}

In this Section, we introduce the simulation cases selected to demonstrate the ability of the CD-ROM approach to improve the performance of POD-Galerkin models. We first present two flow problems to illustrate the benefits of the CD-ROM architecture in the context of fluid mechanics. The first case is the standard configuration of the flow over a cylinder, often used as a benchmark for reduction methods. The second case is the fluidic pinball flow, introduced in \cite{noack2017pinball} for the development of new control strategies. Finally, we introduce the case of the 1D Kuramoto-Sivashinsky which we use to demonstrate the ability of the CD-ROM approach to extend to parametric simulation problems.

\subsection{Flow over a cylinder} \label{Config_cyl}

The two-dimensional incompressible flow over a cylinder has been extensively studied in the context of reduced order modeling \cite{Deane1991pod, Noack2003cylinder} and model identification \cite{brunton2016sindy, loiseau2018constrained} which makes this test case a good initial benchmark for the proposed correction method. The flow is simulated at a Reynolds number of $Re = 100$ based on the cylinder diameter and the velocity of the incoming flow. In this regime, the flow is laminar and exhibits vortex shedding in the wake of the cylinder.

\begin{figure}
    \centering
    \includegraphics[height=0.2\textheight]{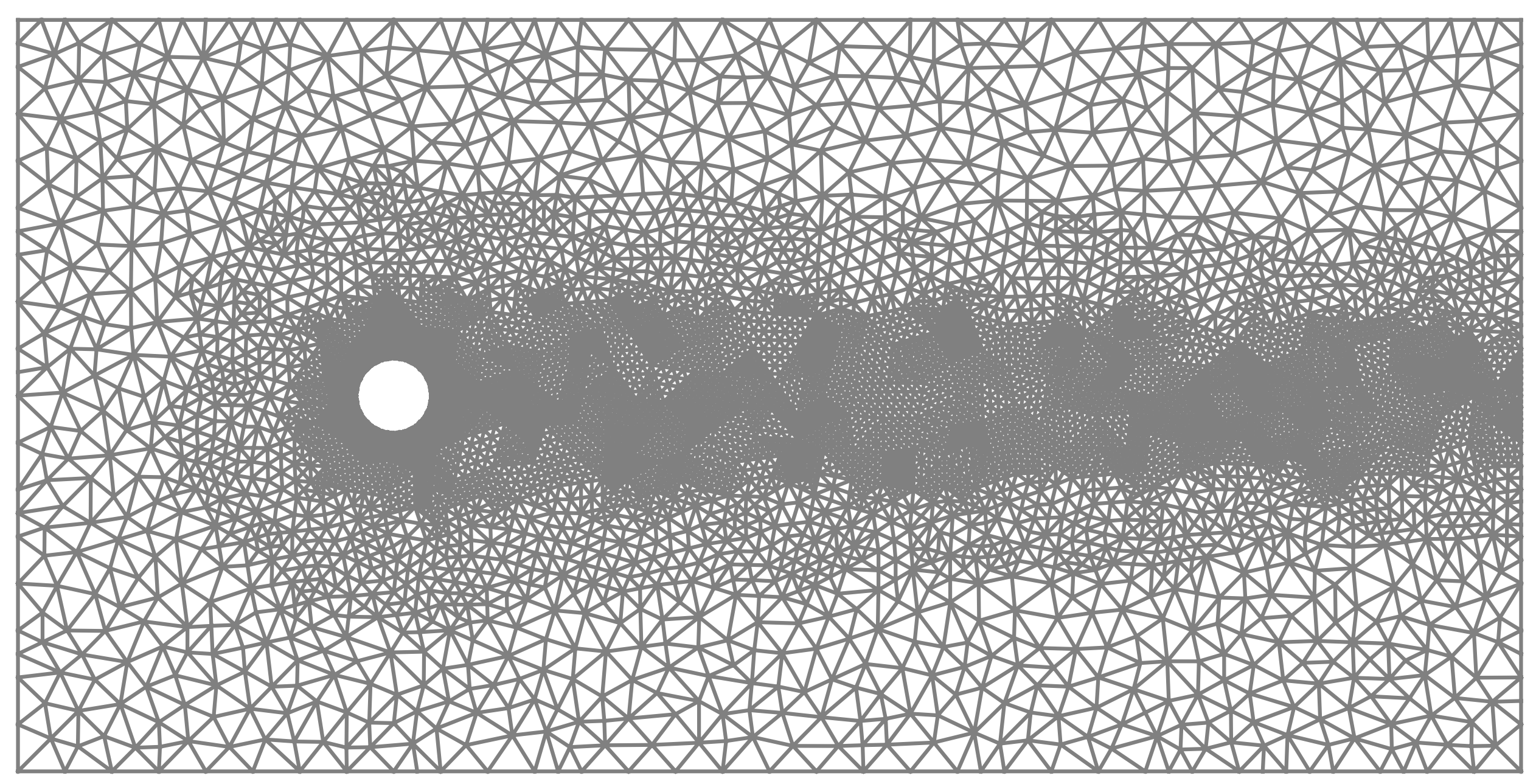}
    \caption{Computational domain for the cylinder case.}
    \label{fig:CylinderMesh}
\end{figure}

The flow is governed by the incompressible Navier-Stokes equations, here solved using the FEniCs finite elements solver \cite{logg2012fenics,alnaes2015fenics}.
The retained mesh is shown \EMmodif{}{in} Figure \ref{fig:CylinderMesh}. It is rectangle-shaped, spanning from $x=-5$ to $x=15$ in the streamwise direction, and from $y=-5$ to $y=5$ in the \EMmodif{}{transverse} direction. The cylinder has a diameter $D=1$, centered around the origin. The inflow is modeled as a uniform axial flow ($\mbf{u}(-5,y)=[U_\infty,0]$) and a free-slip condition is used for the lower and upper boundaries of the rectangle while a no-slip condition is enforced at the cylinder surface. Finally, a stress-free condition is used for the outlet.  

\begin{figure}
    \centering
    \includegraphics[width=0.95\textwidth]{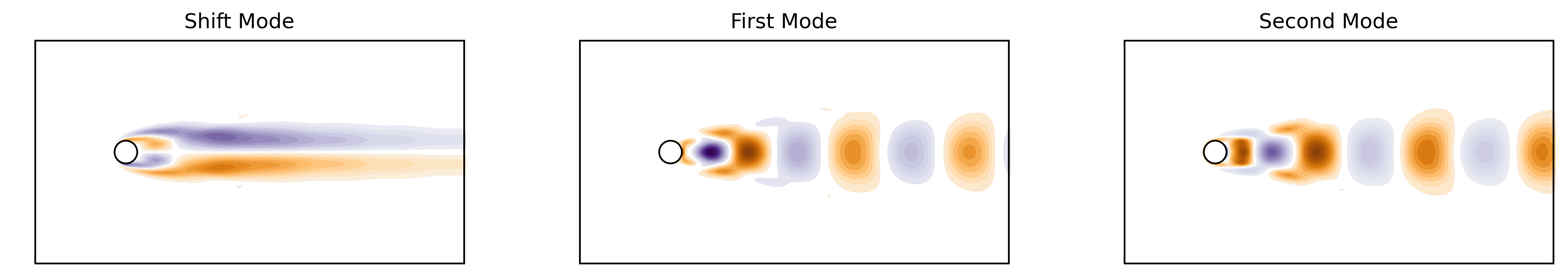}
    \caption{Vorticity fields of the modes selected for the reduced order modeling of the cylinder flow.}
    \label{fig:CylinderBasis}
\end{figure}

A reduced order modeling strategy is employed to obtain a baseline model for the correction approach. The main results are summarised here but we refer to \cite{Noack2003cylinder} for details on the reduction strategy. The vortex shedding regime of the cylinder flow is simulated with the FEniCs solver to obtain snapshot data and compute the POD modes. The first two modes, accounting for more than 95\% of the Frobenius norm of the snapshot matrix are selected. The steady solution of the system is computed with a Newton method to construct a so-called shift mode ($\mbf{a}_\Delta$). This mode is computed as a vector orthogonal to the plane described by the first two modes, pointing to the base flow solution $\mbf{u}_b$, and serves as a \textit{support} for the simulation of the transition of the system from its steady state to the vortex shedding limit cycle. A three-dimensional POD basis is thus finally obtained. They are shown in Fig.~\ref{fig:CylinderBasis} in terms of the vorticity field.

The training data corresponds to the simulation of the transition of the system to the limit cycle of oscillations, starting from an initial condition $\mbf{u}_0$. This initial condition is taken as a point close to the base flow $\mbf{u}_b$ which is the fixed point of the incompressible Navier Stokes equations. Following the procedure of Loiseau \textit{et al.} (\citep{loiseau2018constrained}), we chose
\begin{equation}\label{CylinderU0}
    \mbf{u}_0 = \mbf{u}_b + \epsilon \mbf{v}_1,
\end{equation}
where $\mbf{v}_1$ is the first POD mode, and $\epsilon > 0$ is a small coefficient used to perturb the unstable base flow. 

\begin{figure}
    \centering
    \includegraphics[width=\textwidth]{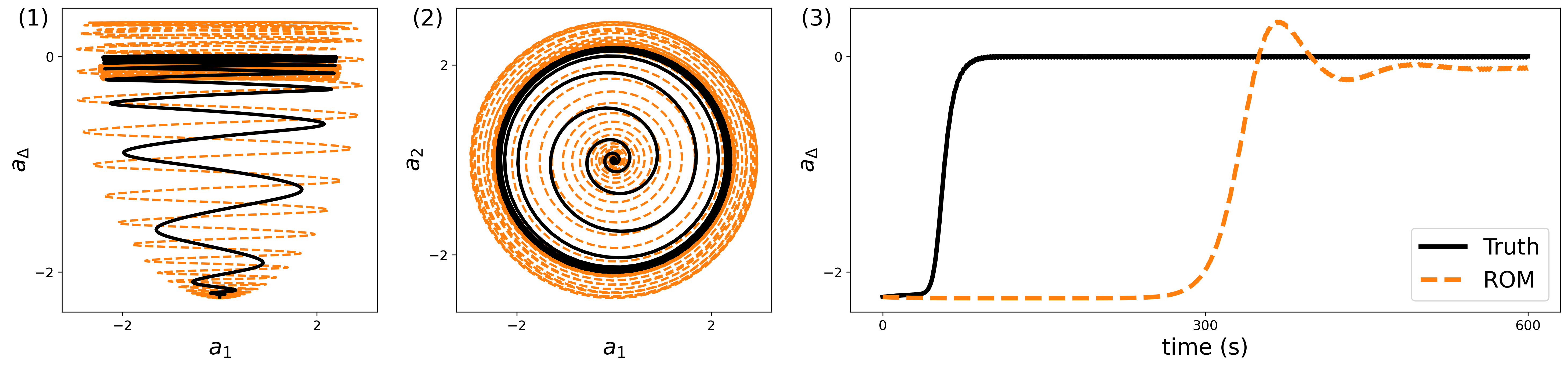}
    \caption{\EMmodif{}{True trajectory and simulation of an uncorrected Galerkin ROM.} Plots (1) \& (2) show the phase space trajectories projected on the $\mbf{a}_1-\mbf{a}_\Delta$ and $\mbf{a}_1-\mbf{a}_2$ planes, while (3) shows the time evolution of the shift mode coefficient.}
    \label{fig:UncorrectedCylinder}
\end{figure}

Through Galerkin projection of the discretised Navier-Stokes equations, a system of 3 coupled ODEs is obtained, describing the dynamics of the reduced coordinates vector. The results of the simulation of the transition using both the Finite Elements model and the Galerkin ROM are displayed \EMmodif{}{in} Figure \ref{fig:UncorrectedCylinder}. Even though the three equation model is able to simulate the transient dynamics, its trajectory strongly diverges from the projected snapshot data. The transition starts much later than in the full order simulation, due to a growth rate of the ROM's transition lower than what it should be. Another significant issue with the model is its stabilisation around the limit cycle ($\mbf{a}_\Delta = \boldsymbol{0}$) where an \textit{overshoot} can be observed before the vortex shedding regime is established, which is not observed in the snapshot data. These discrepancies between the two trajectories can be attributed to the ignored residual term in the dynamics, making this model a good baseline for the correction approach illustrated in Section \ref{sec:Results}.

A test trajectory is also simulated in FEniCs by taking a random initial condition in the phase space spanned by the 3 selected modes, such that:

\begin{equation}
    \mbf{u}_{0,\textit{test}} = \sum_{i=0}^3 \mbf{v_i} a_i,\qquad a_i \sim \mathcal{N}(0,1),
\end{equation}
where $\mbf{v}_i$ are the POD modes, and $a_i$ are random reduced coordinates sampled from a normal distribution. Starting from this initial condition, the finite element model is simulated in FEniCs for $700$ \EMmodif{}{seconds} to ensure the system reaches the oscillation regime. The performance of the various models on this data trajectory are presented in section \ref{sec:Results}.

\subsection{Fluidic pinball}\label{subsec:PinballPres}

\begin{figure}
    \centering
    \includegraphics[height=0.21\textheight]{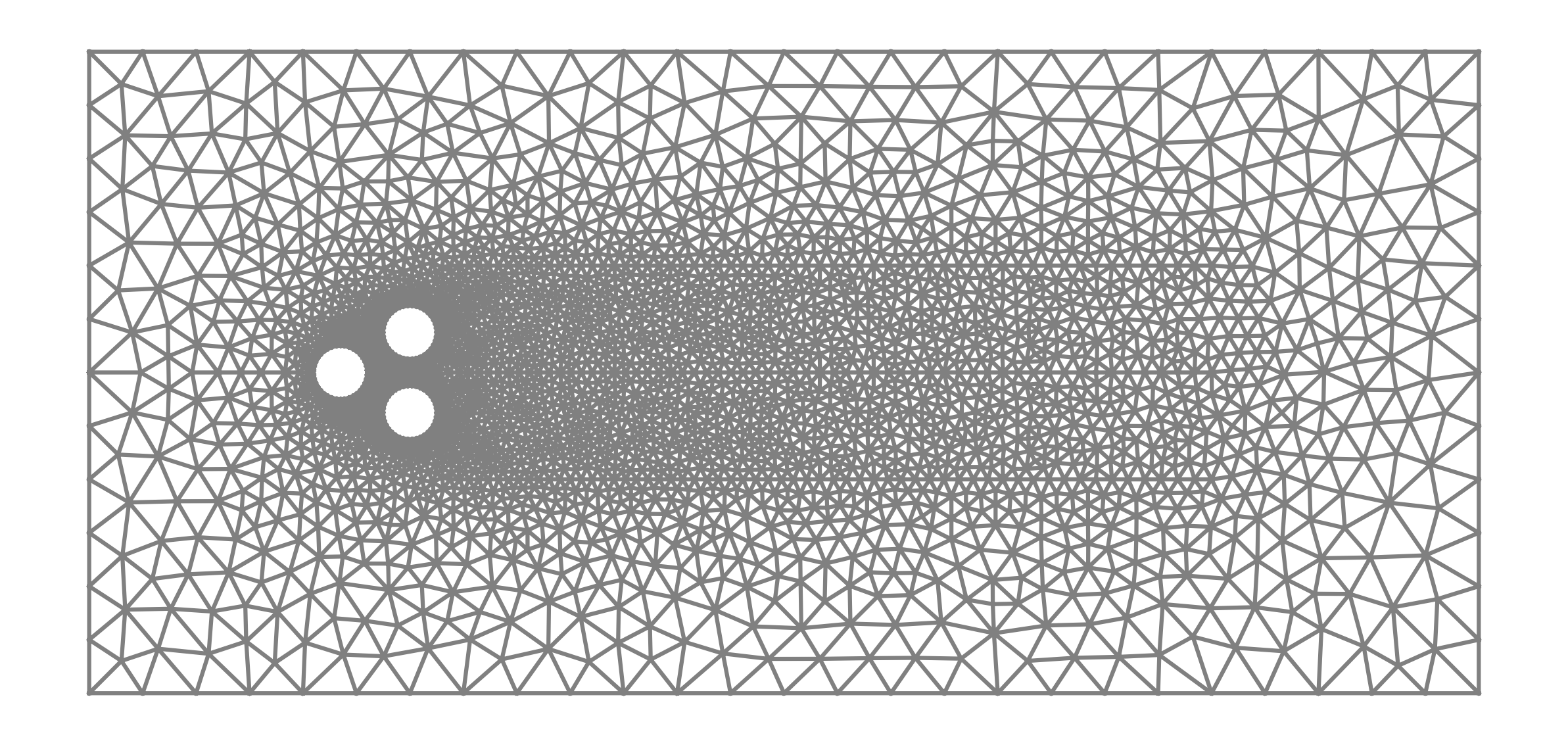}
    \caption{Computational domain of the fluidic pinball case.}
    \label{fig:PinballMesh}
\end{figure}

The second case used to demonstrate the approach is the so-called fluidic pinball. Initially proposed as a challenging test bed for the development of control laws \cite{noack2017pinball}, the fluidic pinball case offers a good trade-off between complexity of its dynamics and interpretability \cite{deng2019pinball,deng2021chaos}. The simulation domain (Figure \ref{fig:PinballMesh}) is composed of three equidistant cylinders, each generating vortices in the wake which interact to create rich dynamics. The mesh used for the simulation was provided by the authors of \cite{cornejomaceda2018pinball}. Displayed \EMmodif{}{in} Figure \ref{fig:PinballMesh}, the domain is a rectangle spanning from $x=-6$ to $x=20$ in the streamwise direction, and $y=-6$ to $y=6$ in the transverse direction. Three identical cylinders with diameter $D=1$ are arranged in an equilateral triangle, with centers' coordinates $(0,0.75),(0,-0.75)$ and $(-1.25,0)$ respectively. The boundary conditions are identical to those of the cylinder case in Sec.~\ref{Config_cyl}. The inflow is modeled as a uniform axial flow, the upper and lower boundaries of the computational domain are modeled as free-slip, while a no-slip condition is used for the walls of the three cylinders and the outlet is modeled as stress-free.

The flow is simulated at a Reynolds number of $Re=130$. At this Reynolds number, the flow dynamics have been shown to be chaotic \cite{deng2021chaos}, which makes it a challenging problem for any correction model as the smallest error on the dynamics will make the simulated trajectory diverge exponentially fast in time from the truth. \EMmodif{}{These} kind of chaotic problems are starting to get traction as interesting benchmarks for forecasting tasks and modeling problems \cite{gilpin2021chaos}. The flow is simulated for $1800$ \EMmodif{}{seconds} in the chaotic regime, which, based on the ergodic property of the system, yields a trajectory long enough to be representative of it's attractor. 

\begin{figure}
    \centering
    \includegraphics[width=0.8\textwidth]{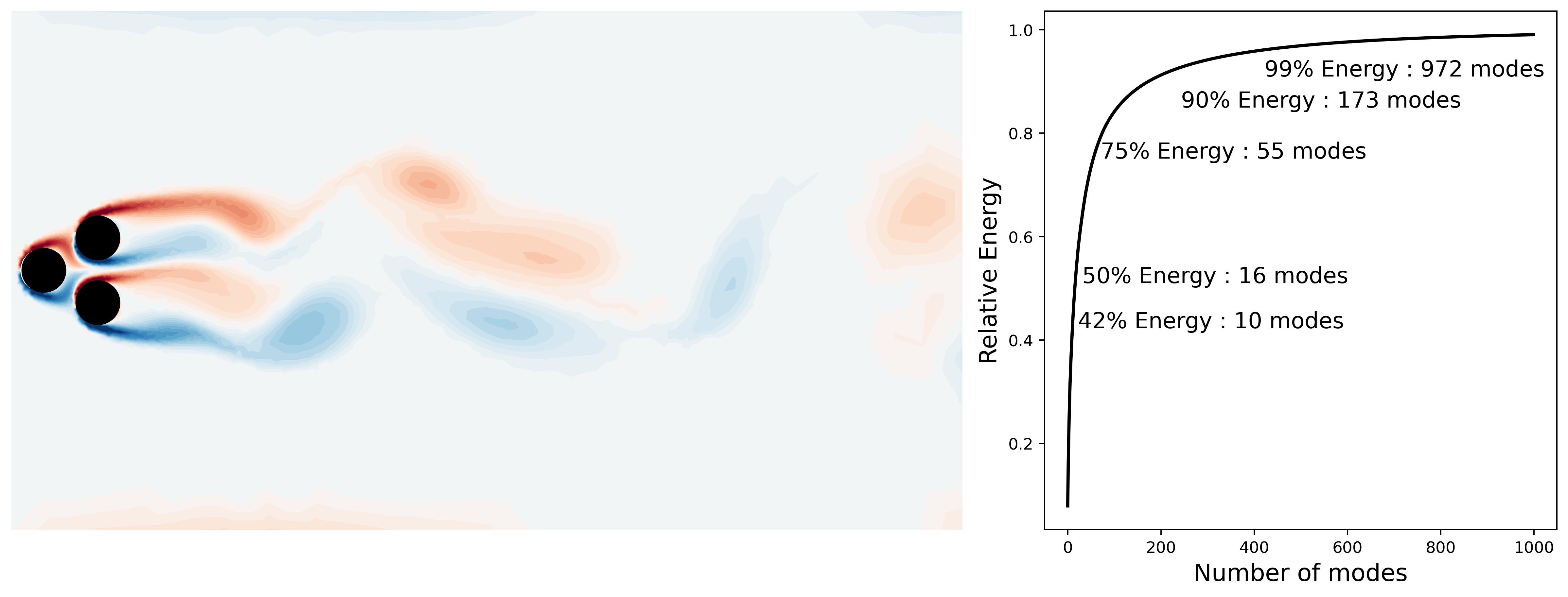}
    \caption{Flow complexity of the pinball case. Left: Vorticity field of the flow in the chaotic regime. Right: Spectrum of the snapshot matrix.}
    \label{fig:PinballIntro}
\end{figure}

It should be noted that the pinball flow is much more complex than the cylinder, as evidenced from Figure \ref{fig:PinballIntro} where it is seen that many POD modes are required to account for a significant part of the energy. Almost a thousand modes would be required to capture 99\% of the Frobenius norm of the snapshot data, while only 8 are required to achieve the same accuracy in the cylinder case.

\begin{figure}
    \centering
    \includegraphics[width=0.95\textwidth]{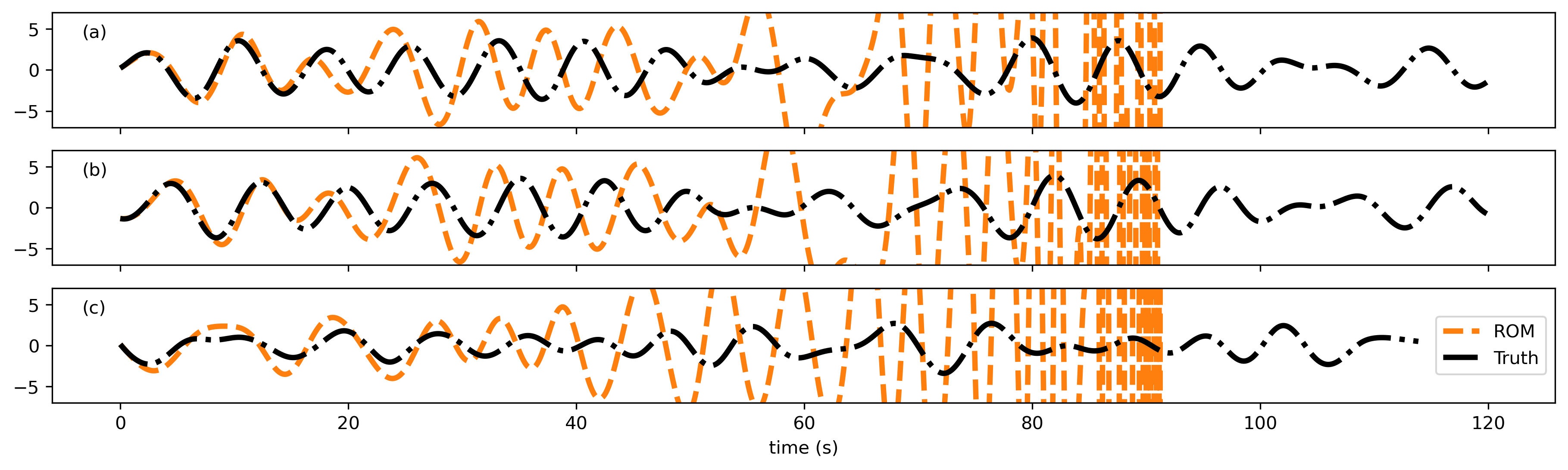}
    \caption{From top to bottom: time evolution of the first 3 POD coefficients. \textit{Black line}: true value of the pinball's POD coefficients, \textit{orange line}: trajectories obtained by integrating an \EMmodif{}{uncorrected} Galerkin reduced order model constructed using the first 10 POD modes.}
    \label{fig:UncorrectedPinball}
\end{figure}

We chose to build a POD-Galerkin model of this flow using only the first 10 POD modes. Although this choice is somewhat arbitrary, it was made to challenge the correction method, as the mean reconstruction error of about $60\%$ has an important impact on the approximated dynamics. Indeed, the obtained reduced model quickly separates from the original trajectory, as expected from a chaotic system. More problematic is the fact that the reduced model is very unstable and diverges after $~70 $\EMmodif{}{s}  of simulation, as shown \EMmodif{}{in} Figure \ref{fig:UncorrectedPinball}. Application of the correction method is presented in Section \ref{sec:Results}, a discussion on the impact of the number of modes is provided in \ref{Annex:NrModes}, while additional information on the computational cost of the approach is given in \ref{Annex:CompCost}.

\subsection{Parametric Kuramoto-Sivashinsky equation}\label{subsec:KS_Intro}

Finally, to illustrate the ability of the proposed method to extend to parametric problems, we introduce the case of the 1D parametric Kuramoto-Sivashinsky (KS) equation. This case is often used to validate physical modeling methods as it is fairly inexpensive to simulate and presents non linear dynamics. Moreover, depending on the parameters used for the simulation, the dynamics become chaotic, making it significantly more challenging to forecasting approaches. The KS equation is formulated as follows:

\begin{align}
    & \frac{\partial u}{\partial t} = -\frac{1}{2}\nabla \cdot u^2 - \Delta u - \nu \Delta^2 u,\label{eq:KS}\\
    & u(x+L,t) = u(x,t),\label{eq:periodcondition}\\
    & u(x,0) = g(x),
\end{align}
where $L$ is the length of the 1D simulation domain, $g$ is the initial condition and $\nu$ is a parameter that controls the degree of dissipativity of the system. Taking inspiration from \cite{wang2020recurrent}, we propose to learn a corrected ROM for this problem under varying $\nu$ values. As in \cite{wang2020recurrent}, we chose $L=22$ and $g$ is computed as the sum of the four leading Fourier modes with coefficients $0.06$. The problem is discretised spatially on a basis of $N = 513$ Fourier modes, and integrated in time through the semi-implicit third order scheme from \cite{kar2006sirk3}. This choice of discretisation implies that the periodicity condition (equation \ref{eq:periodcondition}) is satisfied by construction. The simulation is carried out for a duration of $50$ seconds, using a time step ($\Delta_t$) of $0.025 s$.

\begin{figure}
    \centering
    \includegraphics[width=\textwidth]{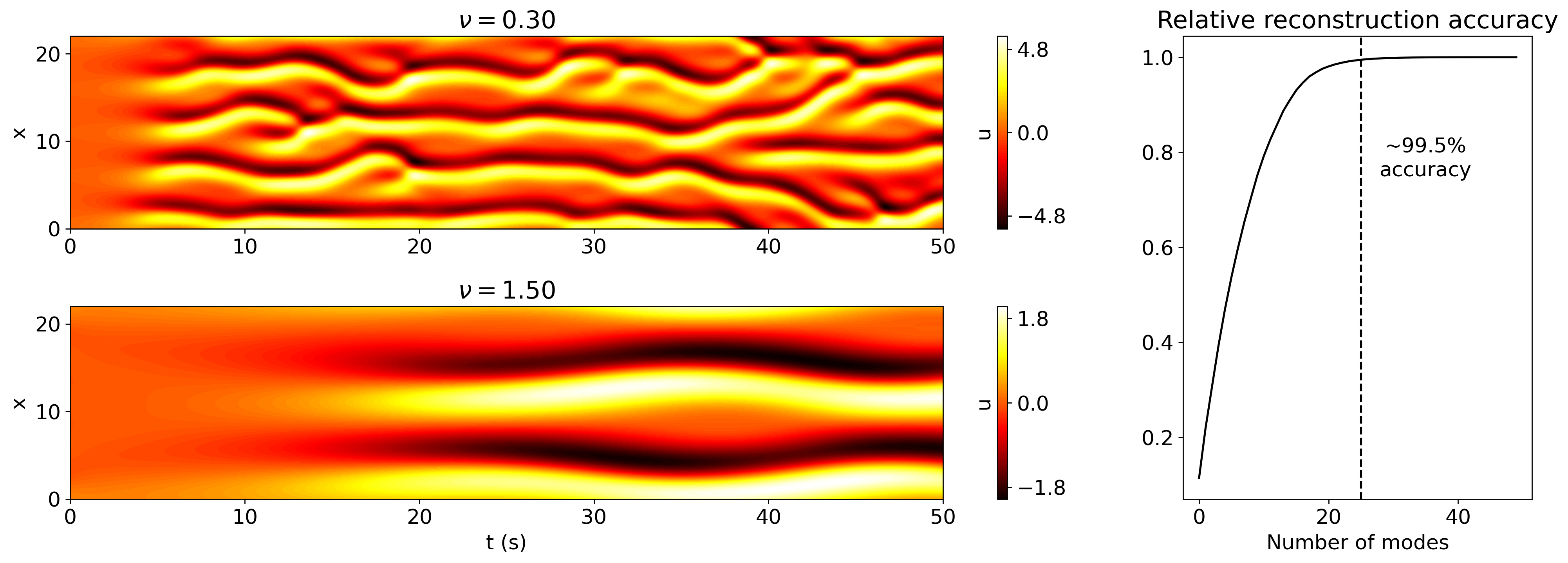}
    \caption{\textit{Left}: Simulations of the KS equation carried out under two different parameter values. \textit{Right}: Relative snapshot reconstruction accuracy against the number of POD modes used for reconstruction.}
    \label{fig:KS_Intro}
\end{figure}

As in \cite{wang2020recurrent}, the parameter $\nu$ is varied in the range $[0.3,1.5]$. As mentioned in the previous paragraph, this parameter controls the degree of dissipation in the system, thus, low $\nu$ values lead to more chaotic dynamics and a harder model reduction task. This is represented on figure \ref{fig:KS_Intro}, which displays the differences between simulations carried out at the limits of the chosen parameter range. 

To create a reduced model of the system, the simulation problem is solved for 25 parameter values selected within a log-linear range from $\nu=0.3$ to $\nu=1.5$. With this initial data, the proper orthogonal decomposition of the snapshot matrix is computed. The evolution of the relative snapshot reconstruction error depending on the number of selected modes is shown \EMmodif{}{in} Figure \ref{fig:KS_Intro}. To assemble the Galerkin reduced model, we select the 25 leading POD modes, which account for more than $99.5\%$ of the information in the snapshot data. The computed POD modes form a basis $V$, which can be used to approximate the solution field $\mbf{u}(t,\nu)$ computed for a given time and parameter value as follows: $\mbf{u}(t,\nu) \approx \tilde{\mbf{u}}(t,\nu) = V\mbf{a}(t,\nu)$.

As in the non-parametric case, computing the vector of reduced coordinates $\mbf{a}(t,\nu)$ is sufficient to fully determine the approximate solution $\tilde{\mbf{u}}(t,\nu)$. Finally, the Galerkin projection method described in section \ref{sec:ModelReduction} is applied to the KS equation, yielding the following reduced model:

\begin{align}\label{eq:KS_Reduced}
    &\dv{\mbf{a}}{t} = -\frac{1}{2} \mbf{a}^\transpose \tilde{\mathcal{Q}} \mbf{a} - 
    \tilde{\mathcal{L}} \mbf{a} - \nu \tilde{\mathcal{L}}^2 \mbf{a},\\
    &\tilde{\mathcal{Q}}_{i,j,k} = <v_i,\nabla (v_j v_k)>, \qquad i,j,k = 1,...,r\\
    &\tilde{\mathcal{L}}_{i,j} = <v_i, \Delta v_j>, \qquad i,j = 1,...,r\\
    &\tilde{\mathcal{L}}^2_{i,j} = <v_i, \Delta^2 v_j>, \qquad i,j = 1,...,r
\end{align}

where $v_i$ are the POD modes, and $<\cdot,\cdot>$ is an inner product defined over the computational domain. It can be noted that the $\tilde{\mathcal{Q}}$ and $\tilde{\mathcal{L}}$ operators are the one dimensional equivalent of the reduced Navier-Stokes operators introduced in section \ref{sec:ModelReduction}. Similarly, the operator $\tilde{\mathcal{L}}^2$ is a linear operator corresponding to the fourth order derivative in equation \ref{eq:KS}. To test the model, 62 test parameter values are selected randomly in the range $\nu \in [0.3,1.5]$ following a log-uniform distribution. 

\begin{figure}
    \centering
    \includegraphics[width=\textwidth]{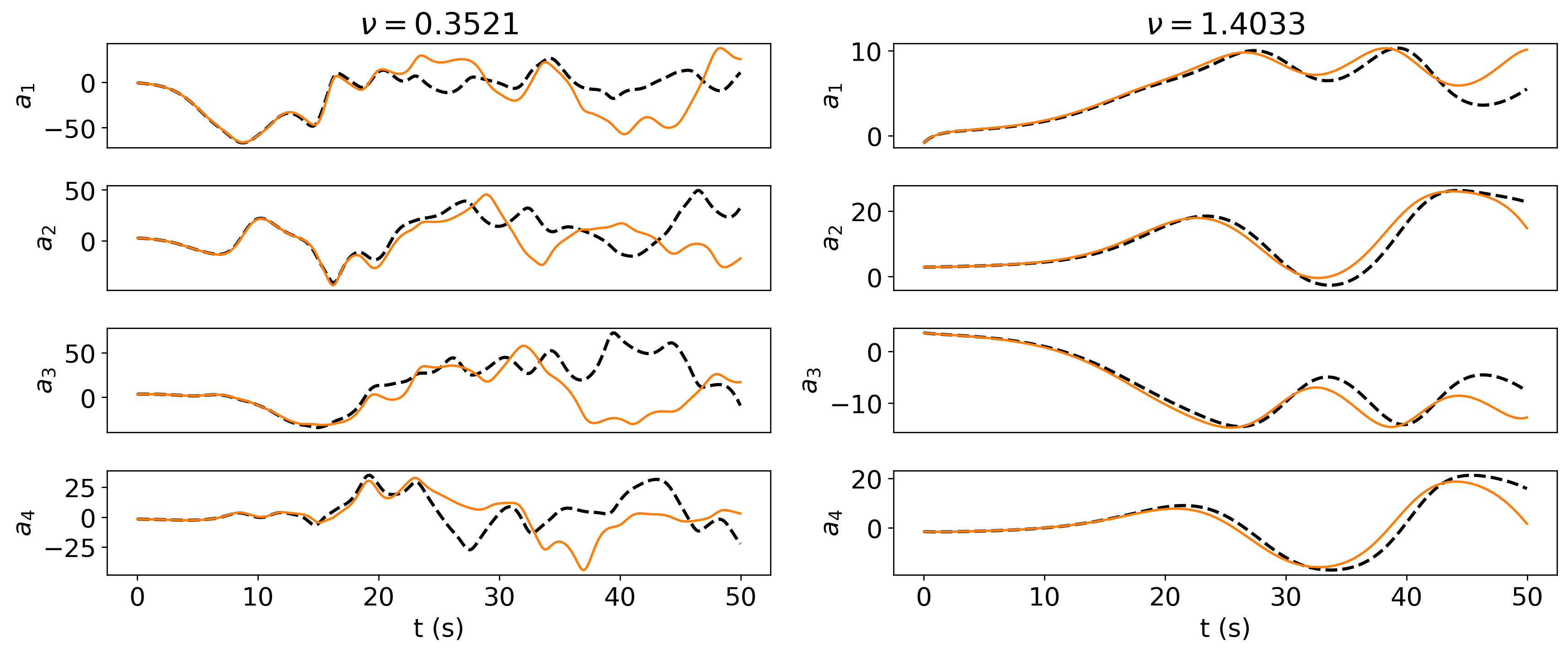}
    \caption{Coefficients of the first 4 POD modes simulated with the uncorrected Galerkin ROM under different parameter values. \textit{Dashed line}: Projected DNS data, \textit{full line}: Simulation of \EMmodif{}{uncorrected} the Galerkin ROM.}
    \label{fig:KS_Uncorrected}
\end{figure}

The uncorrected Galerkin ROM is simulated using the semi-implicit scheme from \cite{kar2006sirk3}. Figure \ref{fig:KS_Uncorrected} presents the results obtained by simulating the Galerkin model under different parameter values. The figure clearly underlines the difficulty of modeling lower $\nu$ values, as we observe that the Galerkin model diverges quickly from the true trajectory. Results obtained by augmenting the Galerkin ROM with the CD-ROM architecture are presented in section 7.

\section{Results and discussion}\label{sec:Results}

\EMmodif{}{In this section, we present the results obtained by applying the CD-ROM method to the imperfect Galerkin ROMs presented in the previous section. While some of the design choices regarding each specific cases are discussed in the following paragraphs, we refer the reader to \ref{sec:TrainingStrat} for a description of the various training details and hyper-parameter choices.} 

\subsection{Cylinder case} 

The reduced model derived in Section \ref{sec:CasePres} above was shown to be efficient for the simulation of the vortex shedding regime, but not suited to the simulation of transient dynamics. To apply the proposed correction procedure to this model, the modeling terms introduced in Sec.~\ref{sec:DataDrivenResidual} are added to the reduced model. The dimension of the memory is chosen to be ten times the dimension of the reduced state. The residual $\mathcal{R}$ and encoder $\mbf{E}$ are defined as multi layer perceptrons, using the Rectified Linear Unit activation function. Finally, the diagonal of the memory matrix $\mbf{\Lambda}$ is initialised at random from a normal distribution. The model is trained with the Adam optimizer, and the trajectory loss introduced earlier   (Eq.~\eqref{eq:trajectory_loss}) is used in combination with additional regularization terms. Details on the regularization of the loss are provided in \ref{sec:Regularization}. 

\begin{figure}
    \centering
    \includegraphics[width=\textwidth]{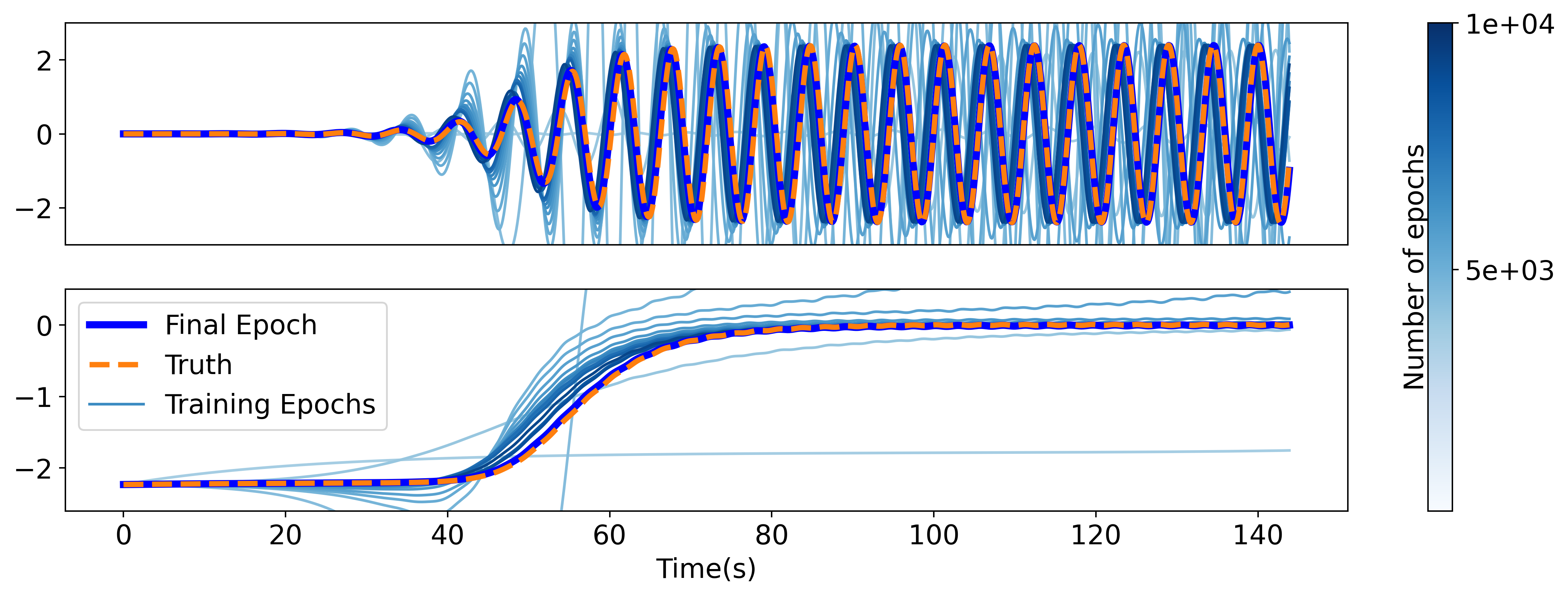}
    \caption{CD-ROM trajectory obtained at different levels of training. The first mode's amplitude is displayed \EMmodif{}{in} the top panel, while the third mode's amplitude is shown \EMmodif{}{in} the bottom panel.}
    \label{fig:CylinderTrain}
\end{figure}

As described in section \ref{sec:CasePres}, the model is trained on the trajectory data obtained by simulating the transition from an initial condition close to the base flow. Since this target trajectory starts close to the base flow of the system, which is stationary, the initial memory can be computed with minimal error through the following integral:

\begin{align*}
    \mbf{y}(0) &= \mbf{E}(\mbf{a}^\star_b) \int_{-\tau_{\mathrm{min}}}^{0} e^{ \mbf{\Lambda}  \tau} \ddroit\tau
\end{align*}
where $\mbf{a}^\star_b = \mbf{V}_r^\transpose \mbf{u}_b$ are the reduced coordinates of the base flow and $\tau_{\mathrm{min}}$ is the longest time horizon defined by the $\mbf{\Lambda}$ matrix as in Equation \eqref{Taumin}. Finally, the parameters of the models are progressively optimised to reproduce the true transition trajectory as shown \EMmodif{}{in} Figure \ref{fig:CylinderTrain}. 

\begin{figure}
    \centering
    \includegraphics[width=\textwidth]{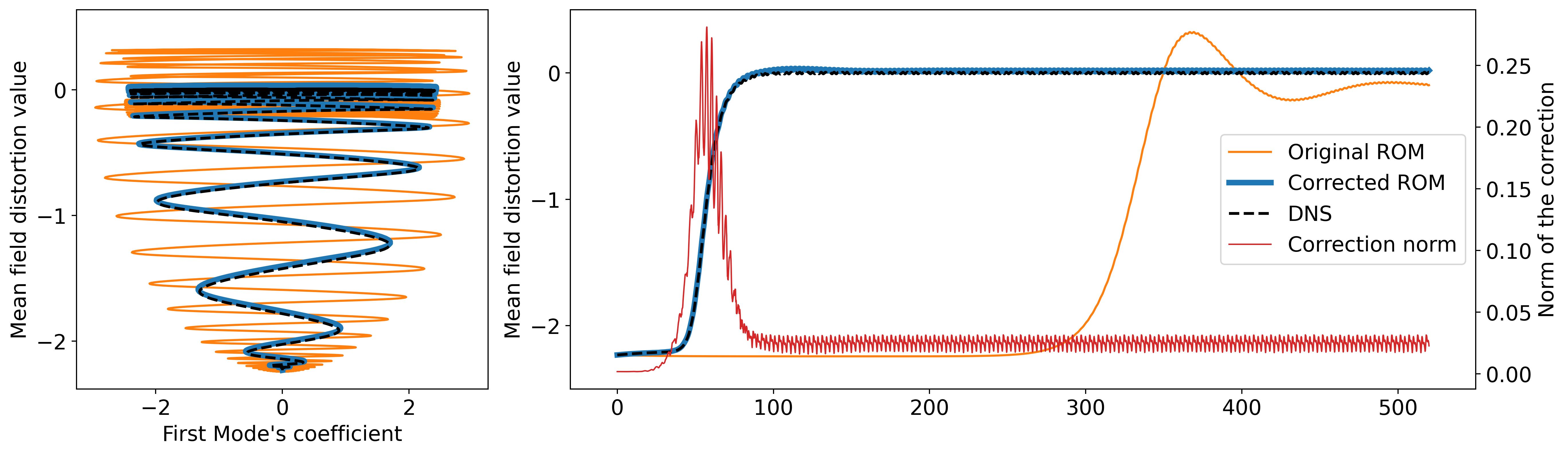}
    \caption{Results obtained with the corrected ROM for the simulation of the cylinder flow's transient dynamics. Left: phase space trajectory from base flow to limit cycle simulated with the finite element solver, the original Galerkin ROM and our corrected model. Right: time evolution of the shift mode's coefficient simulated with the same models, as well as the norm of the correction applied by our model.}
    \label{fig:CylinderNODE}
\end{figure}

 The CD-ROM model is integrated in time using an adaptive RK-45 scheme. Simulation results on the training trajectory are presented on Figure \ref{fig:CylinderNODE}. It can be seen that the corrected model follows the training trajectory perfectly, triggering the transition at the right time, and correcting the oscillations of the original ROM during the stabilisation on the limit cycle. Moreover, the graph shows that the correction applied by the residual model is strong during the transition, where the original ROM struggles, and becomes minimal during the rest of the trajectory.

\begin{figure}
    \centering
    \includegraphics[width=\textwidth]{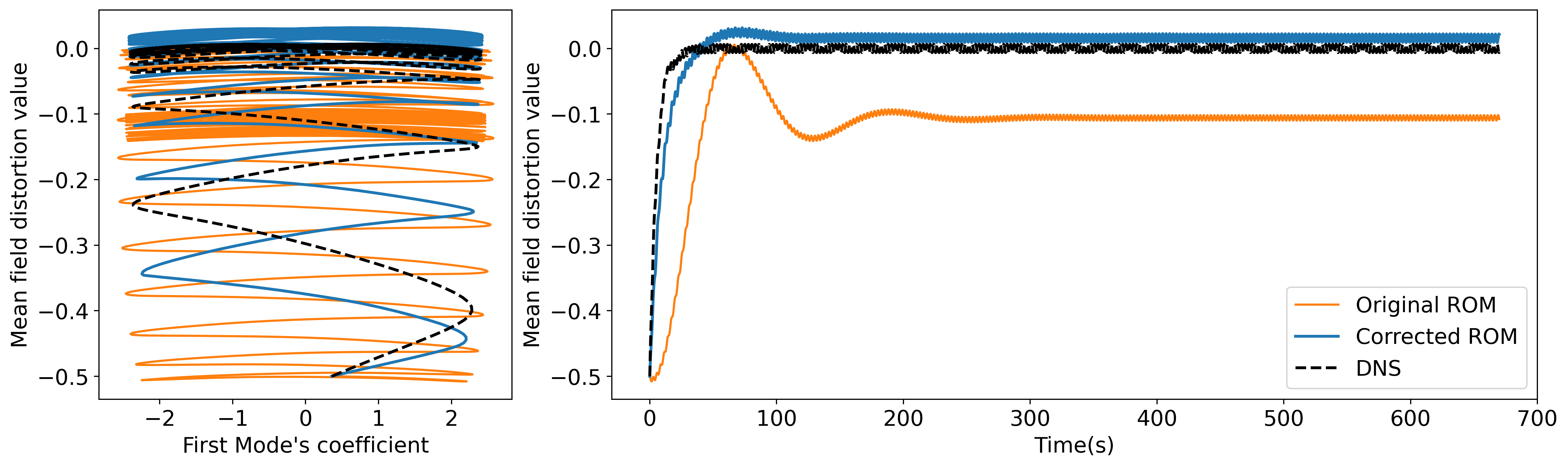}
    \caption{Simulation results obtained on a test trajectory. As in Figure \ref{fig:CylinderNODE}, the first graph represents the phase space trajectory, while the time evolution of the shift mode's coefficient is presented on the second graph.}
    \label{fig:TestCylinderNODE}
\end{figure}

Finally, we present results of the performance of the model on the test trajectory. The first 8 \EMmodif{}{seconds} of DNS simulation are used to initialise the memory following Equation \eqref{eq:memInit}. Figure \ref{fig:TestCylinderNODE} presents the performance of each model on this trajectory. Because the starting point is not close to the base flow, the uncorrected ROM instantly exhibits transient dynamics, however, the growth rate of the transition is still inaccurate and the shift mode's trajectory presents the same non-physical oscillations around the limit cycle observed in the training trajectory. The corrected model does much better than the original ROM, simulating a more accurate transition, and stabilising on the limit cycle almost perfectly. 

\subsection{Fluidic pinball results}

While the cylinder case discussed above offers a simple test bed for the presentation of the approach and its potential, it has already been shown that very parsimonious models could be used to model its dynamics \cite{loiseau2018constrained}, which suggests this configuration might not require a high dimensional neural network to learn a correction term. The case of the fluidic pinball is more challenging and can better underline the ability of our method to handle complex physics. As presented in the previous section, the POD-Galerkin approach is not well suited to the reduction of the pinball case. The number of modes required to reconstruct the snapshot data with a satisfying accuracy is very large and using a small number of modes leads to an unstable model. 

To apply the correction approach to the pinball case, a correction model is built. The encoder and residual models are multi layer perceptrons and the Sigmoid Linear Unit activation function is used as it leads to smoother integration. The weights of the residual model are initialised to be close to 0 so that the ROM is initially almost uncorrected. Diagonal entries of the memory matrix are initialised as a log-linear range of time horizons, ranging from $0.6$ to $3.84$ \EMmodif{}{seconds}. The training data consists of $1800$ \EMmodif{}{seconds} (15000 snapshots) of DNS simulation in the chaotic regime. The leading two thirds of the simulated DNS trajectory are used for training while the remaining third is set aside for testing. 

As presented in the previous section, the uncorrected 10-mode ROM is unstable and diverges after $70$ \EMmodif{}{s} of simulation. As a result, trying to optimise the correction model for long trajectories directly would lead to a very unstable training process. To address this issue and stabilise the training, the corrected ROM is trained on sub-trajectories of only $10$ \EMmodif{}{seconds} This length is chosen as it is short enough for the model to remain stable and long enough for the impact of the encoder on the memory to be accounted for. Once a good correction model has been trained for $10$ \EMmodif{}{seconds} long trajectories, the length is progressively augmented to attain the target horizon of $120$ \EMmodif{}{s}. Besides the stabilisation of the training, using sub-trajectories also allows for parallel training. Multiple sub-trajectories can be sampled from the snapshot data and integrated in parallel on a GPU to dramatically speed-up training, more information about the training strategy as well as the training parameters used is provided in \ref{sec:TrainingStrat}.

\begin{figure}
    \centering
    \includegraphics[width=0.9\textwidth]{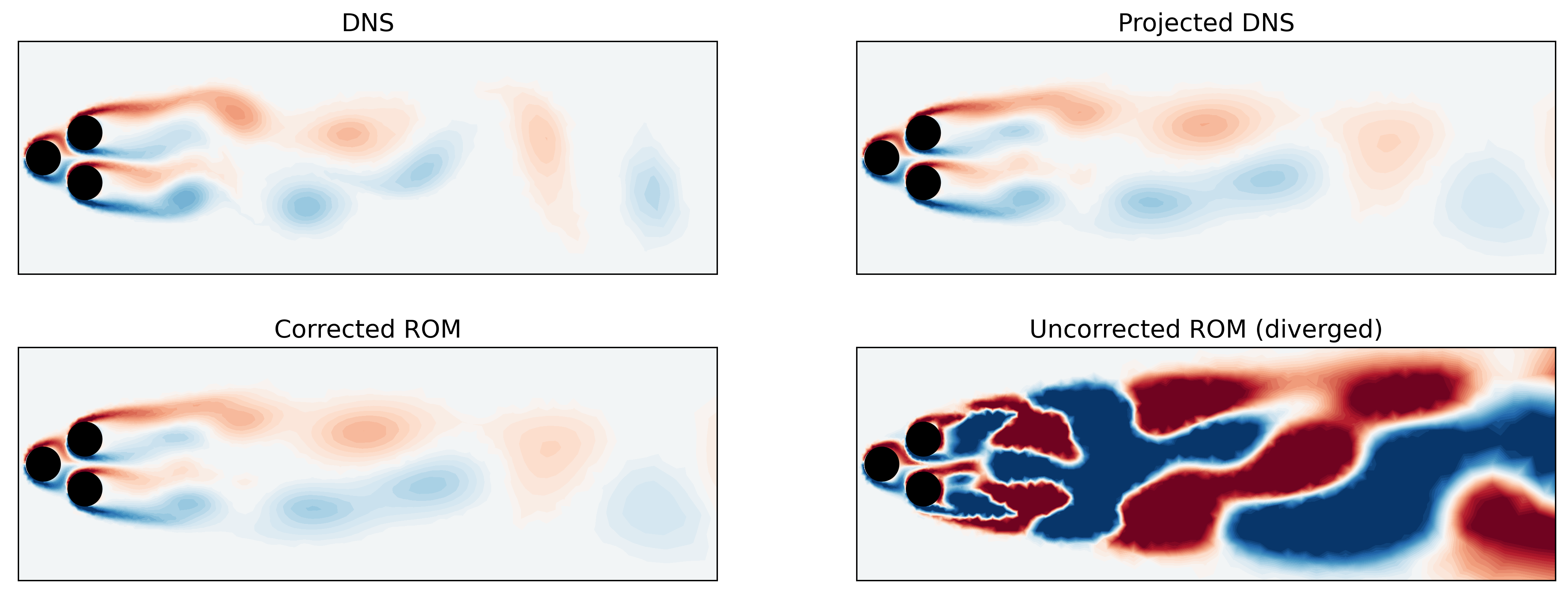}
    \caption{Vorticity fields obtained after 110 \EMmodif{}{s} of simulation in the chaotic regime.}
    \label{fig:VorticityComparisons}
\end{figure}

The pinball correction model was trained to follow true trajectories up to $120$ \EMmodif{}{seconds}. Results of the simulated flow fields are presented on Figure \ref{fig:VorticityComparisons}. It can be seen that the projection on the 10-dimensional POD basis effectively filters part of the spatial structures, leading to the divergence of the uncorrected reduced model. In contrast, the corrected ROM is able to reproduce the projected flow field accurately. 

\begin{figure}
    \centering
    \includegraphics[width=0.99\textwidth]{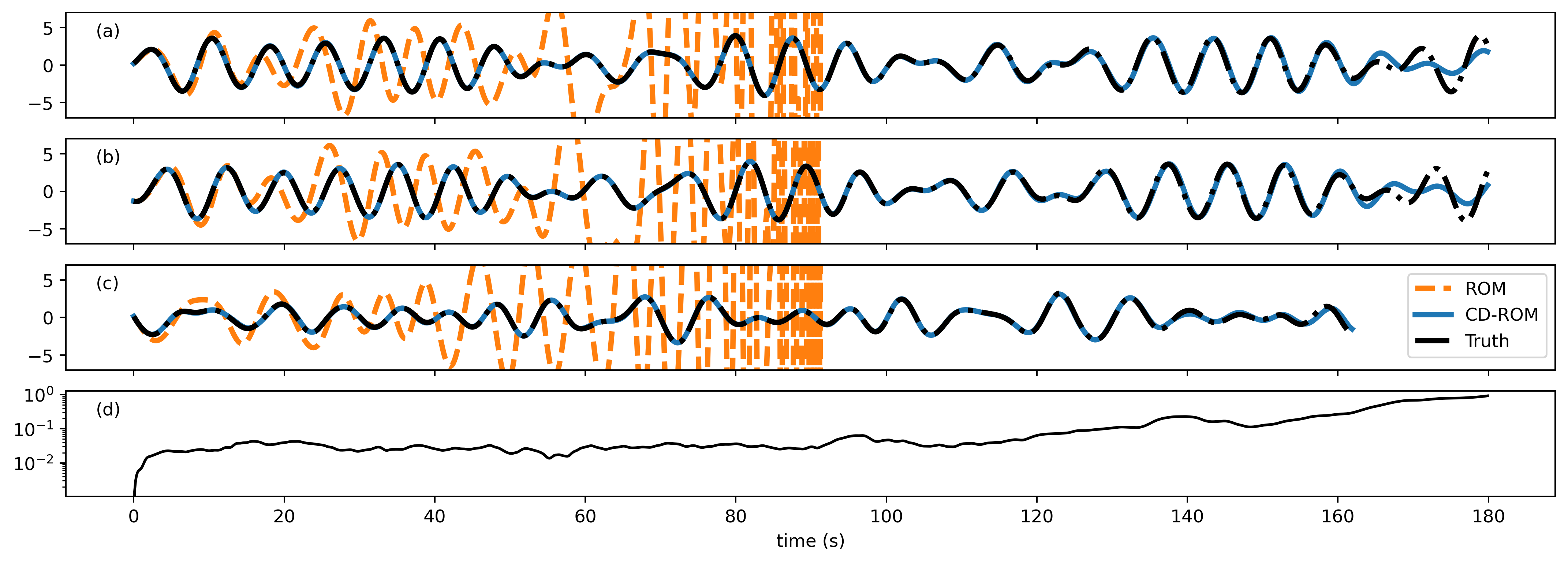}
    \caption{Results of the correction approach applied to the 10-mode ROM of the fluidic pinball. The initial condition is a random point selected in the training trajectory. Plots (a),(b),(c) describe the coefficients of the first three POD modes simulated with different models. Plot (d) presents the relative Euclidean distance between the true data and the trajectory simulated with the CD-ROM.}
    \label{fig:PinballResultsTrain}
\end{figure}

As with the cylinder case, the CD-ROM is simulated with an adaptive RK-45 scheme. Trajectory results simulated from a condition in the training basis are presented on Figure \ref{fig:PinballResultsTrain}. The model starts quickly diverging from the true trajectory after the training horizon (120 \EMmodif{}{s}) as is expected from the chaotic nature of the problem. Deriving a model to perfectly follow the DNS trajectory would here make little modeling sense. 

More interesting is the fact that, despite leaving the training trajectory, the corrected model does not become unstable, even when integrated for over $1000$ \EMmodif{}{seconds} with an initial condition outside of the training data. This suggests that the dynamics correction learned by the model has some physical consistency, dissipating the necessary energy which would otherwise have made the simulation diverge in the uncorrected case.

\begin{figure}
    \centering
    \includegraphics[width=0.95\textwidth]{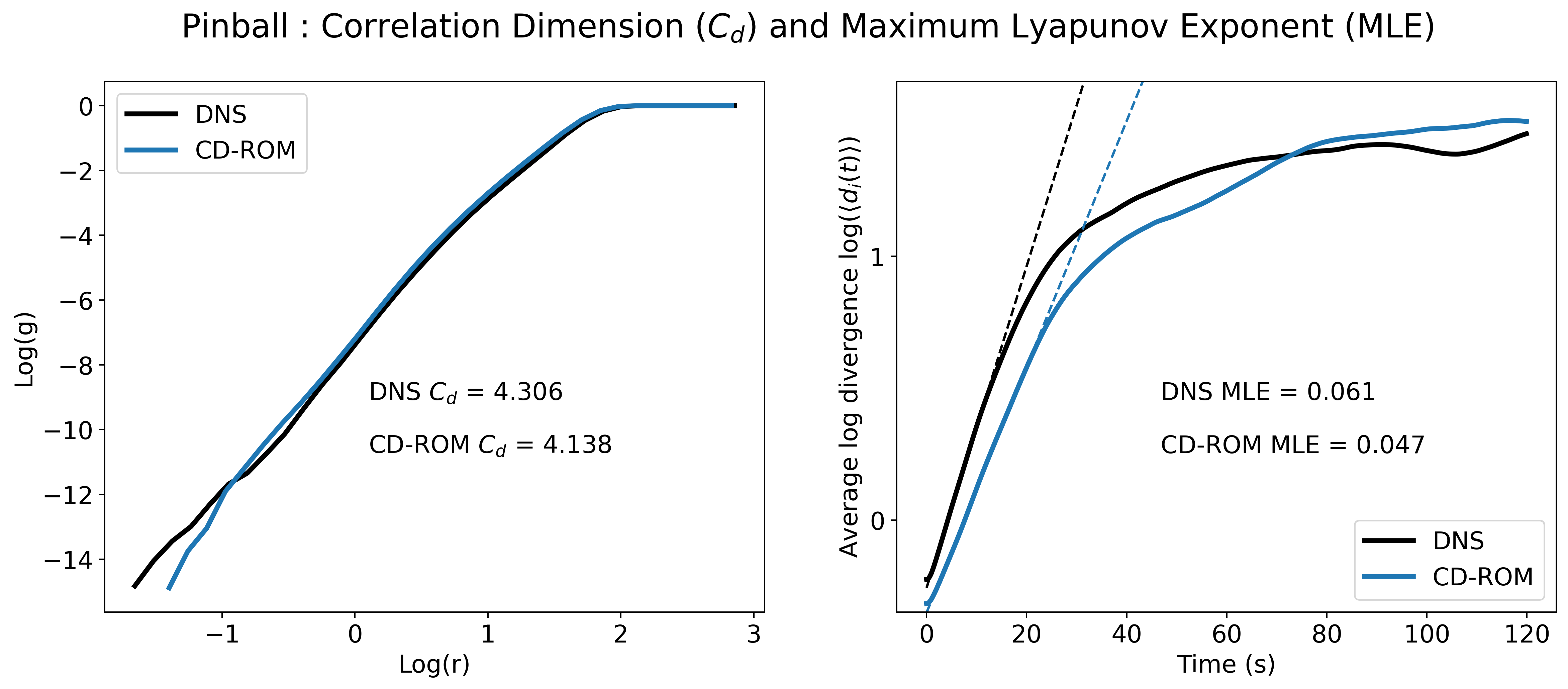}
    \caption{Statistics of the 10-dimensional attractors described by the projected DNS data (black) and the CD-ROM (blue). Left: Average number of points $g$ in a sphere of radius $r$, the data follows the law $g = r^{C_d}$ where $C_d$ corresponds to the correlation dimension of the attractor. Right: Time evolution of the average distance $d$ between trajectories starting from arbitrarily close initial conditions on the attractor. The data follows the law $d = e^{tl_e}$ where $l_e$ corresponds to the maximum lyapunov exponent of the system at hand. }
    \label{fig:PinballStatistics}
\end{figure}

\begin{figure}
    \centering
    \includegraphics[width=0.95\textwidth]{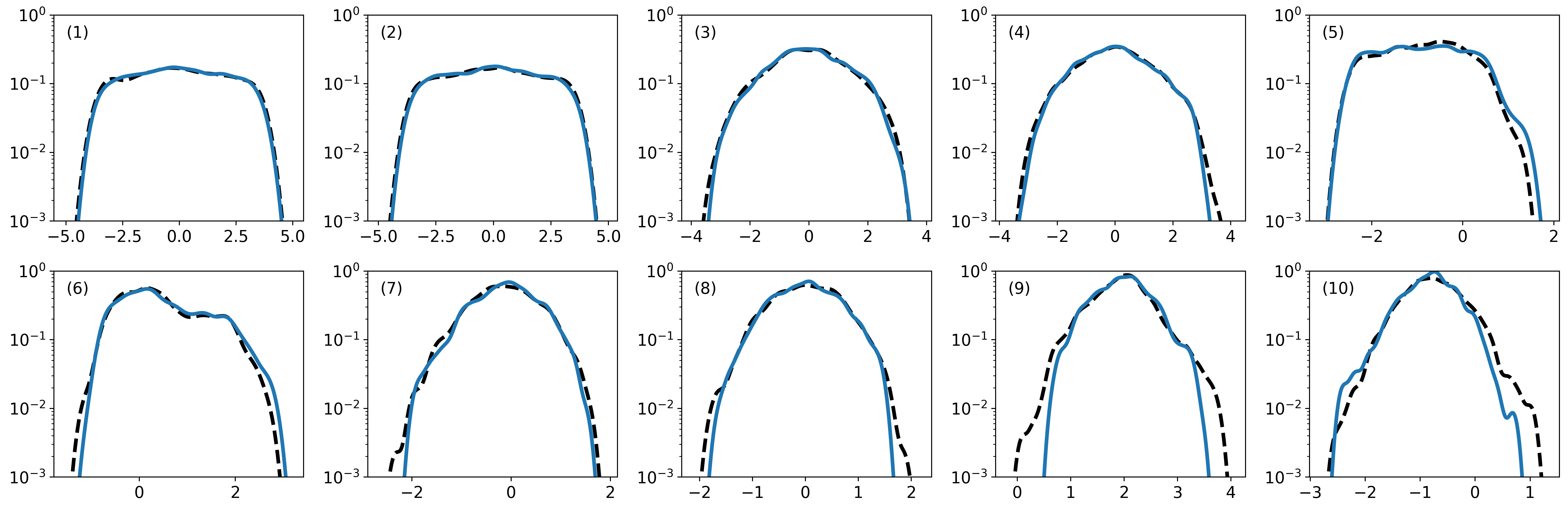}
    \caption{Estimated probability density functions for the coefficient of each mode. \textit{Plain blue line}: Statistics of the trajectory simulated with the corrected model; \textit{dotted black line}: Statistics of the projected DNS data. Labels refer to the mode index.}
    \label{fig:LogPdf}
\end{figure}

The intuition that the model was able to learn a Physics-compatible correction is confirmed when looking at the statistics of the attractor spanned by the CD-ROM's trajectory. Using the \texttt{nolitsa} library \cite{nolitsa}, the correlation dimension (\cite{grassberger1983cd}) was estimated, as well as the maximum Lyapunov exponent (\cite{wolf1985mle}) of the corrected and true trajectories. The results are shown  \EMmodif{}{in} Figure \ref{fig:PinballStatistics} where the model is seen to reproduce well the characteristics of the true attractor. One can also look at the probability distributions of the mode's amplitudes, presented on Figure \ref{fig:LogPdf}. Once again, the simulated trajectory reproduces the results of the true simulation. 

Note that long trajectories (several hundreds of \EMmodif{}{seconds}) were simulated with the corrected model to obtain these statistics. Due to the chaotic nature of the problem and the length of the integration period, the model has visited parts of the attractor different from those seen in the training trajectory. Despite this, the CD-ROM remains stable and describes an attractor with statistics that are very similar to those of the true attractor, further supporting the approach for the reduced modeling of complex dynamics.

\subsection{Parametric KS equation results}

In this paragraph, we present the results obtained by applying the CD-ROM architecture to a parametric case, the KS equation presented in section \ref{subsec:KS_Intro}. To do so, the reduced Galerkin model of equation \ref{eq:KS_Reduced} is augmented with the proposed CD-ROM architecture (equation \ref{eq:DeepROM}). To account for the parametric nature of the problem, the coefficient $\nu$ is passed to both the residual ($\mathcal{R}$) and encoder ($\mathbf{E}$) models, yielding the following CD-ROM system:

\begin{equation}
    \begin{array}{c l c c c}
    \dv{}{t}\mbf{a} &=& -\frac{1}{2} \mbf{a}^\transpose \tilde{\mathcal{Q}} \mbf{a} - 
    \tilde{\mathcal{L}} \mbf{a} - \nu \tilde{\mathcal{L}}^2 \mbf{a}& +& \mathcal{R}(\mbf{y}, \nu;\boldsymbol{\theta}_\mathcal{R}), \\
    \dv{}{t}\mbf{\mbf{y}} &=& \mbf{E(\mbf{a}, \nu;\boldsymbol{\theta}_E}) &-& \mbf{\Lambda} \, \mbf{y}.
\end{array}
\end{equation}

The residual and encoder models are both expressed as multi layer perceptrons, using the SiLU activation function. The weights of both neural networks, as well as the memory matrix $\mbf{\Lambda}$, are optimised using the Adam optimizer. As in the previous fluidic pinball case, we start by optimising the model on small sub-trajectories, then gradually increase the length of the sub-trajectories as the model reaches the desired accuracy. The model is trained on the data generated to computed the POD modes and assemble the Galerkin ROM (see Section \ref{subsec:KS_Intro}). This training data corresponds to simulations carried out under 25 different parameter values in the range [0.3,1.5]. After training, the model is tested on 62 new simulations carried out under different parameter values selected randomly following a log-uniform distribution, as described in Section \ref{subsec:KS_Intro}. 

\begin{figure}
    \centering
    \includegraphics[width=\textwidth]{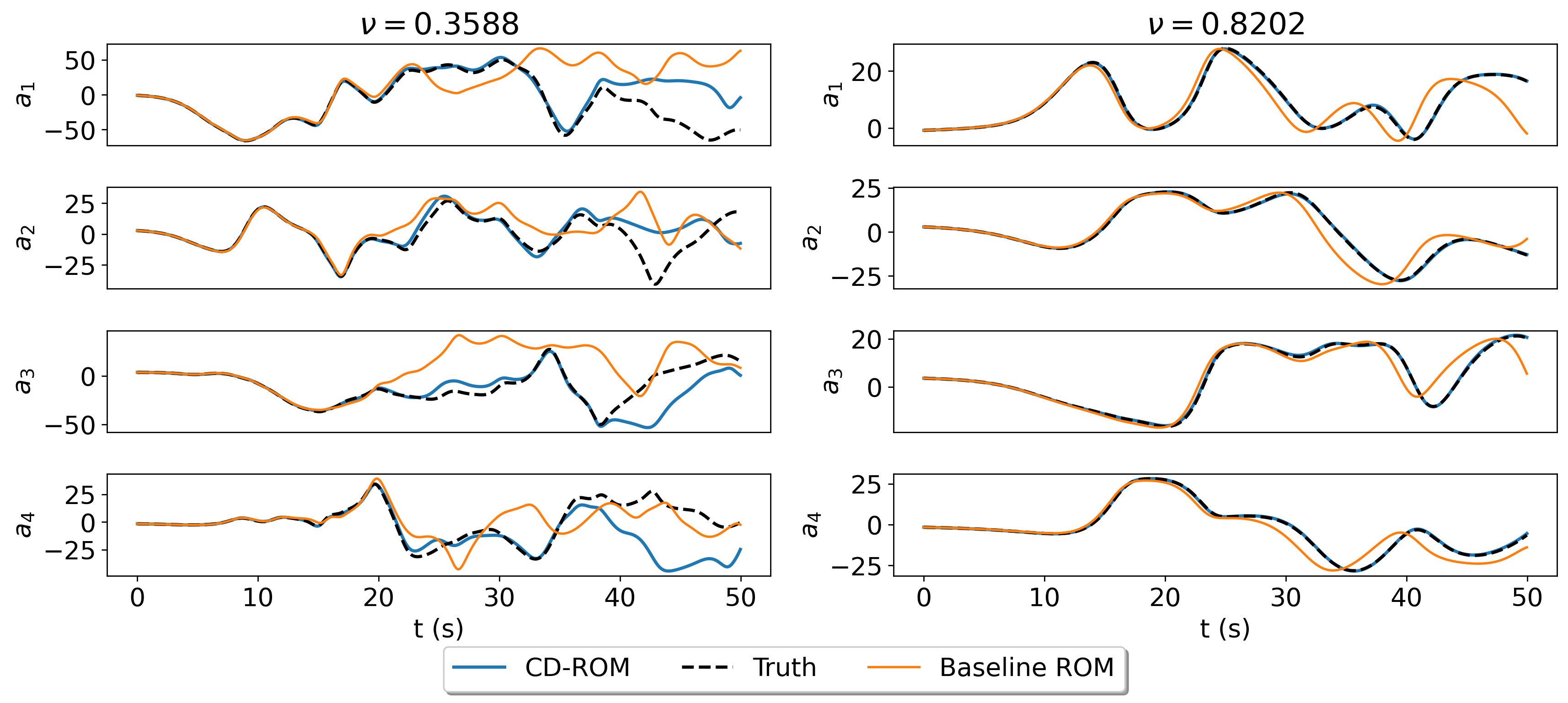}
    \caption{Coefficients of the first 4 POD modes simulated with the CD-ROM as well as the uncorrected Galerkin ROM under different parameter values in the test dataset. \textit{Dashed line}: Projected DNS data, \textit{full orange line}: Simulation of the Galerkin ROM, \textit{full blue line}: Simulation of the CD-ROM.}
    \label{fig:KS_Corrected}
\end{figure}

Contrary to the two previous flow cases, it is more efficient to simulate the KS equations using the semi implicit time-stepping scheme of \cite{kar2006sirk3}, thus we use this scheme to integrate the CD-ROM model in time. Figure \ref{fig:KS_Corrected} presents the results obtained by simulating the CD-ROM using $\nu$ values not included in the training data. On the two cases presented, the CD-ROM performs better than the Baseline Galerkin ROM. It can be seen that the model diverges from the true trajectory data earlier in the case where the value of $\nu$  is lower. This is expected as we showed in section \ref{subsec:KS_Intro} that lower $\nu$ values lead to more complex dynamics. 

To assess the performance of the CD-ROM over the whole test set, the Euclidean distance between the simulated reduced coordinates vector $\mbf{a}$ and the projected DNS data $\mbf{a}^\star$ is computed over time, and for every test parameter values:

\begin{equation}
    \label{eq:KS_metric}
    d(t,\nu) = \Vert \mbf{a}(t,\nu) - \mbf{a}^\star (t,\nu) \Vert_2.
\end{equation}

\begin{figure}
    \centering
    \includegraphics[width=\textwidth]{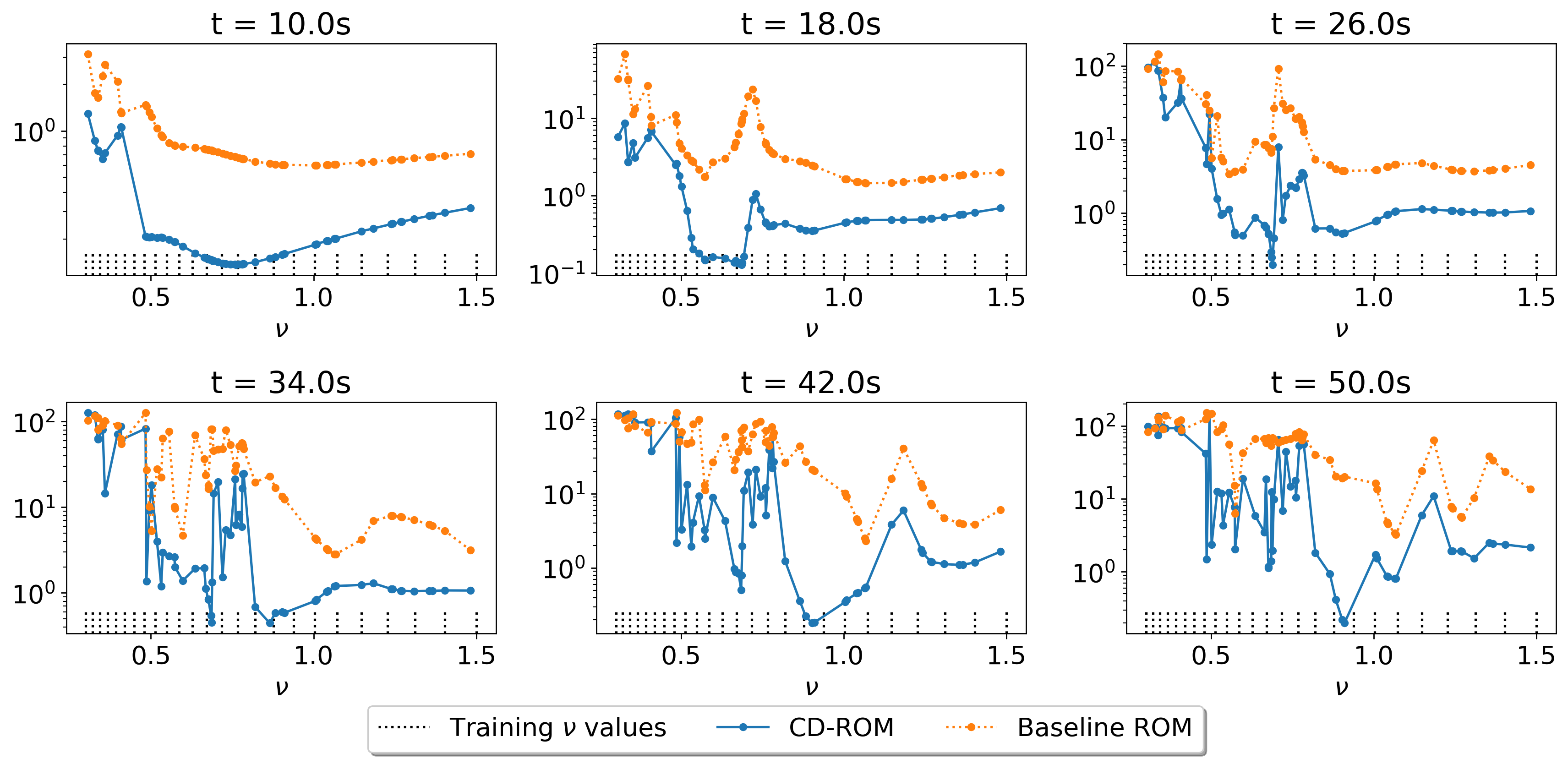}
    \caption{Error metric (equation \ref{eq:KS_metric}) computed for each test parameter value.}
    \label{fig:KS_Test_Results}
\end{figure}

Figure \ref{fig:KS_Test_Results} presents the values of the error metric (equation \ref{eq:KS_metric}) for every test parameter value at select time steps. The figure shows that the CD-ROM is able to remain significantly closer to the true trajectory than the baseline Galerkin ROM for almost $20$ seconds. The model then behaves differently depending on the parameter value. Cases in the $\nu \in [0.3,0.5]$ range presenting the more chaotic dynamics quickly diverge from the true trajectory, while the CD-ROM is able to beat the baseline on the rest of the test cases for up to $50$ seconds.

These results demonstrate the ability of the CD-ROM architecture to improve the Galerkin model order reduction approach in a parametric setting. The trained CD-ROM model is able to reproduce the dynamics of the full order system better than its uncorrected counterpart, even when using parameter values different from the training conditions.

\section{Conclusion}\label{sec:Conclusion}

This paper was concerned with an augmented reduced order modeling strategy based on the hybridisation of the classical Galerkin projection method and simple neural networks. By studying the limitations of the Galerkin projection, we established links with the field of partially observed systems, which led us to use past observables of the studied system as a critical ingredient for the correction of Galerkin models. Building on this result, the CD-ROM architecture was proposed to extract and exploit useful information from the system trajectory, by embedding the model with a delay differential equation structure. Moreover, the training strategy based on adjoint optimisation ensures \textit{a-posteriori} performance of the model on the training trajectory.

The CD-ROM approach was demonstrated on two simple CFD test cases, namely the flow over a cylinder and the fluidic pinball. Numerical experiments have shown that the corrected models were able to capture the true dynamics with a high degree of accuracy, reproducing the true transition in the case of the cylinder flow, and following the training trajectory for multiple Lyapunov times in the fluidic pinball case. Moreover, these experiments outlined the reliability of the corrected model as it performed better than its uncorrected counterpart even outside the training conditions. The results obtained with the fluidic pinball are particularly encouraging. We showed that the correction model was able to stabilise the original Galerkin model in a consistent physical manner, as the attractor simulated with the CD-ROM approach presents statistics similar to the original attractor.

The ability of the proposed approach to extend to parametric problems was also demonstrated. The CD-ROM method was applied to the case of the Kuramoto-Sivashinsky equation with varying viscosity. After training the model on a small number of parameter values in a selected range, we showed that the CD-ROM approach improved the performance of the baseline Galerkin ROM over the whole parameter range, even when simulating using parameter values outside of the training data. This is of interest for many real-world situations, e.g. industrial applications where a low cost parametric model is a key-enabler.

On-going work investigates the application of the CD-ROM approach on a more concrete industrial use case, to demonstrate the capabilities of the method on a real-world application. We are also looking into the best way to build parametric models of dynamical systems using the CD-ROM, as recent approaches leveraging hyper-networks to dynamically modify the model weights have shown promising results on data driven forecasting tasks.

\appendix

\newpage

\section{Neural ODEs and Adaptive Checkpoint Adjoint}\label{AdjointAnnex}

\subsection{Adjoint Backpropagation}

The adjoint backpropagation algorithm on which Neural ODEs are based can be used to solve problems of the form\footnote{This specific criterion and constraints choice are well suited to our approach. The adjoint backpropagation algorithm can be used to solve more complex problems; however, we chose to restrict the scope to this formulation to clarify the derivation.}:

\begin{equation}
\begin{aligned}
\min_{\theta} \quad & J(x(T))\\
\textrm{s.t.} \quad & \dot{x} - g(x;\theta) = 0\\
  &x(0) = x_0.    \\
\end{aligned}
\end{equation}

Generally, \EMmodif{}{these} kind of problems are solved with a gradient descent method, which requires evaluation of the following gradient:

\begin{equation}
    \dv{J}{\theta} = \frac{\partial J}{\partial x} \frac{\partial x}{\partial \theta}\Big|_T.
\end{equation}

Because the derivative of $\dot{x}(t) = g(x;\theta)$ is parameterised by $\theta$, there is an implicit relation $x = x(t;\theta)$. The estimation of the sensitivity $\frac{\partial x}{\partial \theta}|_T$ requires the consideration of the impact of $\theta$ on the whole time-integration, from $t=0$ to $t=T$. The adjoint backpropagation is used to avoid computing this term, which is done by evaluating the sensitivity of the following Lagrangian:

\begin{align}
    &\mathcal{L} = J + \int_0^T \mu(t)(\dot{x}(t) - g(x;\theta)) \ddroit t\\
    &\dot{x}(t) = g(x;\theta) \implies \dv{J}{\theta} = \dv{\mathcal{L}}{\theta}.
\end{align}

The vector of Lagrangian multiplier $\mu$ is a function of time. Distributing the product in the integral and integrating the first term by parts leads to an expression where the sensitivity $\frac{\partial x}{\partial \theta}\Big|_T$ can be isolated:

\begin{equation}
    \mathcal{L} = J + \Big[\mu x \Big]_0^T - \int_0^T \dot{\mu}x(t) + \mu(t) g(x;\theta) \ddroit t
\end{equation}

\begin{equation}\label{eq:fullSensitivity}
    \implies \dv{\mathcal{L}}{\theta} = \Big(\frac{\partial J}{\partial x} + \mu(T)\Big) \frac{\partial x}{\partial \theta}\Big|_T - \mu(0) \cancelto{0}{\frac{\partial x}{\partial \theta}\Big|_0} - \int_0^T \Big(\dot{\mu} + \mu(t) \frac{\partial g}{\partial x}\Big)\frac{\partial x}{\partial \theta}\Big|_t + \mu(t)\frac{\partial g}{\partial \theta}\Big|_{x=x(t)} \ddroit t.
\end{equation}

A so-called adjoint equation can be derived to avoid having to solve for the sensitivity $\frac{\partial x}{\partial \theta}$. Enforcing a vanishing variation of the Lagrangian \textit{wrt} $x$ at optimality yields:

\begin{equation}\label{eq:AdjointEq}
\begin{aligned}
\dv{\mu}{t} &= -\mu \frac{\partial g}{\partial x}\Big|_t\\
\mu(T) &= -\frac{\partial J}{\partial x}\Big|_T.\\
\end{aligned}
\end{equation}

Solving the adjoint equation (\ref{eq:AdjointEq}) for the values of $\mu(t)$ leads most of the terms in Equation \eqref{eq:fullSensitivity} to vanish, so that: 

\begin{equation}\label{eq:AdjointGrad}
    \dv{\mathcal{L}}{\theta} = - \int_0^T \mu(t) \frac{\partial g}{\partial \theta}\Big|_{x=x(t)} \ddroit t.
\end{equation}

As mentioned earlier, both this integral and the adjoint equation can be easily evaluated if $\dot{x}(t)$ is approximated by a neural network as the required vector-Jacobian products $\mu \frac{\partial g}{\partial x}$ and $\mu(t) \frac{\partial g}{\partial \theta}$ can be easily computed in a deep learning framework.

\subsection{Adaptive checkpoint adjoint}

The original NeuralODE paper (\cite{chen2019node}) proposed integrating forward in time to obtain the initial condition of the adjoint equation evaluated from $x(T)$, while discarding intermediate values $x(t)$. This choice was made with the goal of reducing the memory footprint of the method. However, discarding the intermediate time-steps $x(t)$ means that they have to be recomputed during backpropagation to evaluate the adjoint ODE (\ref{eq:AdjointEq}), as well as the gradient integral (\ref{eq:AdjointGrad}). This can be done as a single backward in time integration by concatenating the different quantities ($x,\mu,\dv{\mathcal{L}}{\theta}$) in a single state vector:

\begin{align}
    z &= \Big[x,\mu,\dv{\mathcal{L}}{\theta}\Big]\\
    \dv{z}{t} &= \Big[g(x;\theta),-\mu\frac{\partial g}{\partial x},-\mu \frac{\partial g}{\partial \theta}\Big]\\
    z(T) &= \Big[x(T), -\frac{\partial J}{\partial x}\Big|_T, \frac{\partial J}{\partial x} \frac{\partial g}{\partial \theta}\Big|_T\Big].
\end{align}

\begin{figure}
    \centering
    \includegraphics[width=0.9\textwidth]{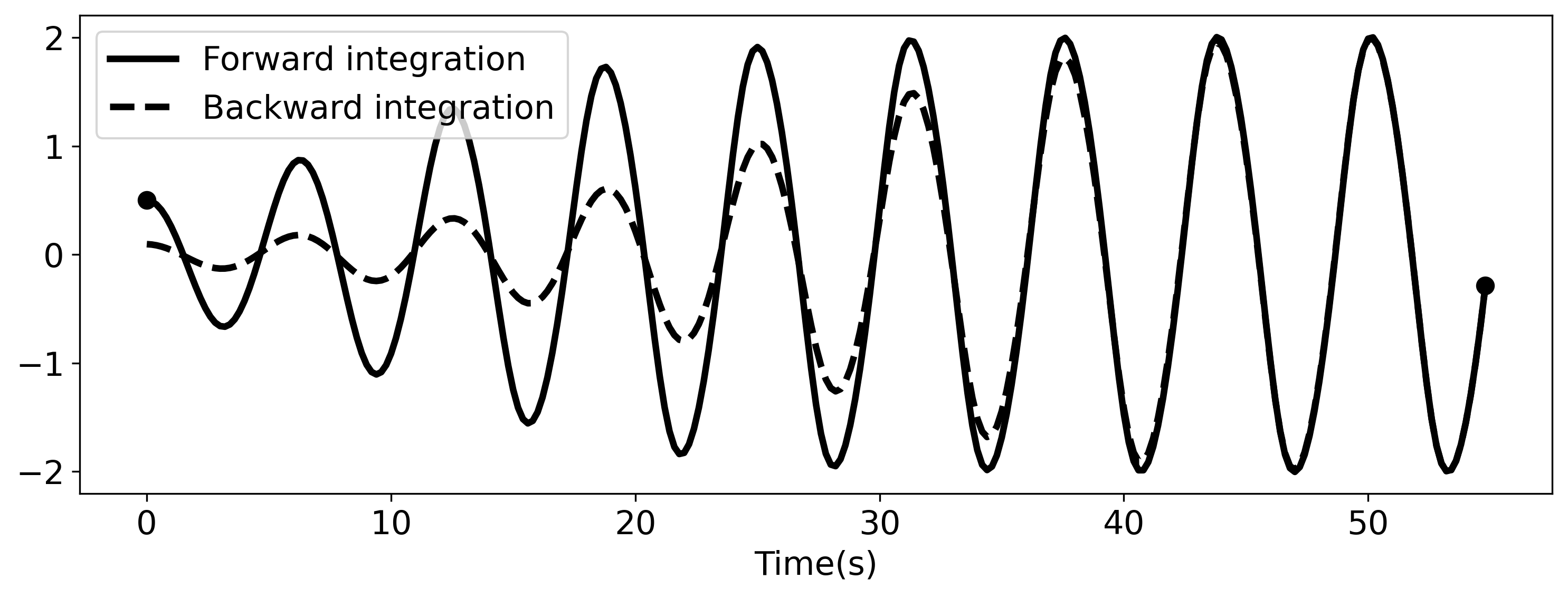}
    \caption{Forward and backward time integration of the Van der Pol oscillator. Numerical errors during the integration lead to the divergence of the two trajectories which should be identical.}
    \label{fig:vanderpol}
\end{figure}

Not only does this increase the computational cost of the method, but it can also lead to erroneous gradients, as numerical errors during integration can lead to differences between the forward and backward trajectories for $x(t)$. This was observed in \cite{zhuang2020aca} and is illustrated on Figure \ref{fig:vanderpol}, which shows that, despite using the same parameters, the forward and backward time trajectories diverge due to numerical errors. To address this issue, one can use the Adaptive Checkpoint Adjoint method. This method retains the intermediate integration steps ($t,x(t)$) and simply evaluates the integrals for the adjoint $\mu$ and the gradient $\dv{J}{\theta}$ at the time steps selected during the forward integration, using the forward trajectory $x(t)$. 

Although this increases the memory footprint of the method, it limits the computational cost of the backward pass, as the time steps are already selected, and the trajectory $x(t)$ does not have to be integrated another time. Most importantly, this limits the numerical errors introduced by the integration schemes, which can have a significant impact on the training, especially in the case of chaotic time-derivatives.

\section{Number of modes for the fluidic pinball reduction}\label{Annex:NrModes}

\subsection{Uncorrected models comparison}

In Section \ref{sec:CasePres}, we present a 10-mode reduced order model of the fluidic pinball problem. The number of modes is chosen somewhat arbitrarily to challenge the method. The more modes are used to model the flow, the better the model will be at reproducing the true dynamics, reducing the complexity of the required residual term. In this Appendix, we provide more insights into the impact of the number of modes on the modeling problem to clarify the choice of using 10 modes to model the pinball flow.

\begin{figure}
    \centering
    \includegraphics[width=0.95\textwidth]{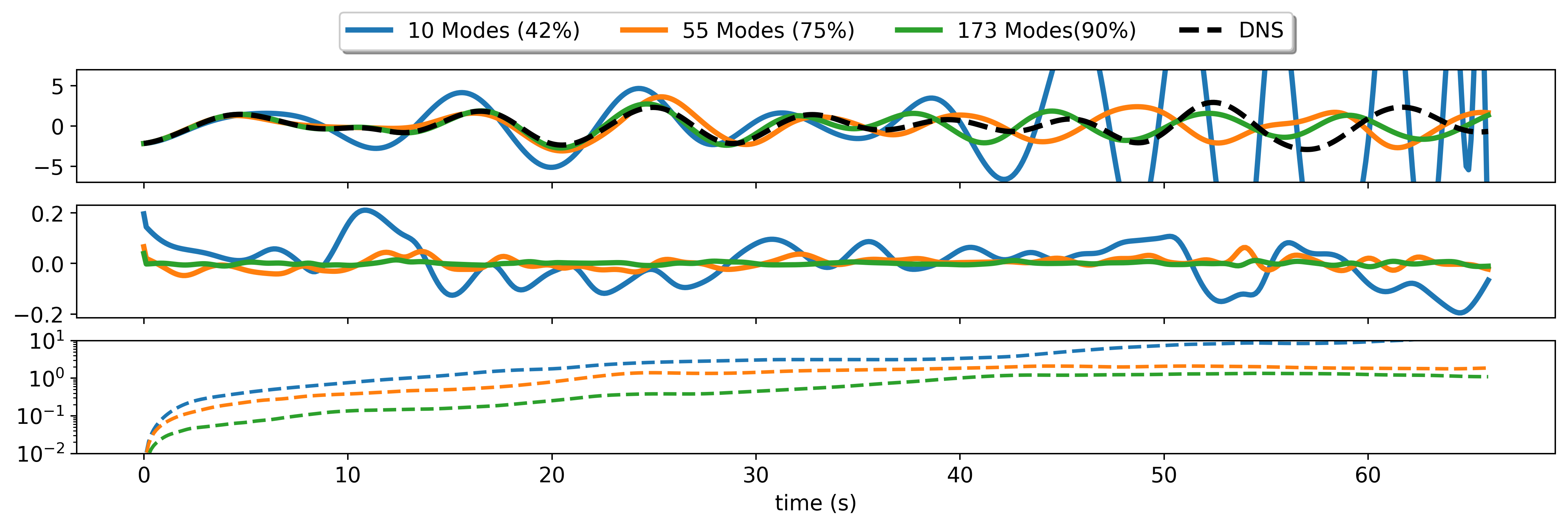}
    \caption{Comparison of the performance of POD-Galerkin models at different degrees of reduction. \textit{Top}: Simulated value for the amplitude of the first POD mode; \textit{center}: Value of the closure term for the first POD mode computed on the training trajectory; \textit{bottom}: Relative distance between simulated and true trajectories.}
    \label{fig:CorrectionComparisons}
\end{figure}

Figure \ref{fig:CorrectionComparisons} presents the performance of different reduced models of the pinball flow. It is clearly seen that increasing the number of modes is beneficial for the performance of the reduced models. The magnitude of the closure term, as well as the speed at which the reduced model diverges from the true trajectory, are reduced when the number of modes increases. This Figure also shows that a higher number of modes leads to more stable reduced models. However, further experiments showed that even well resolved models such as the one using 173 modes were not stable and would diverge in certain conditions. 

\subsection{Training convergence}

To add to the argument, a comparison of the training convergence between two models using different number of modes is discussed. Two models are built, using respectively 10 and 55 POD modes, and the correction architecture is trained using the same parameters, presented in Table \ref{tab:CompareTrainingParameters}.

\begin{center}
\begin{tabular}{|c|c|}
\hline
    Memory Dimension &  5 $\times$ POD dimension\\
    Residual lr & $10^{-3}$\\
    Encoder lr & $10^{-3}$\\
    $\Lambda$ lr & $2 \, 10^{-4}$\\
    Optimiser & AdamW \\
    \hline
\end{tabular}
\captionof{table}{Parameters, e.g. learning rates (lr), used for the training of the compared CD-ROM architectures.}\label{tab:CompareTrainingParameters}
\end{center}

In both cases, the same loss is used, combining the optimisation of the distance between simulated and true trajectory with the residual regularization introduced in \ref{sec:Regularization}:

\begin{equation}\label{eq:CompareLoss}
    J = \frac{1}{r} \left( \frac{1}{n_t}\sum_{i=1}^{n_t} \left\| \mbf{a}(i  \Delta_t) - \mbf{a}^\star(i \Delta_t)\right\|^2_2 + c \frac{1}{n_t} \sum_{i=1}^{n_t} \left\|  \mathcal{R}(\mbf{y}(i \Delta_t)) - \mbf{\dot{a}^\star}(i \Delta_t)\right\|_2^2 \right).
\end{equation}
where $c = 0.1$ is a constant weighting the importance of the stochastic residual regularization term \textit{w.r.t.} the trajectory loss. Notice that, to ease the comparison between the two models, the loss is scaled by the number of modes used in the ROM. The two models are trained in the same fashion, sub-trajectories of a hundred time steps are sampled in the training base and the loss (\ref{eq:CompareLoss}) is optimised until a chosen threshold ($5 \times 10^{-4}$) is reached, at which point the length of the sub-trajectories is increased by fifty time steps. This process is repeated until the model is able to reproduce sub-trajectories of a thousand time steps. 

\begin{figure}
    \centering
    \includegraphics[width=0.95\textwidth]{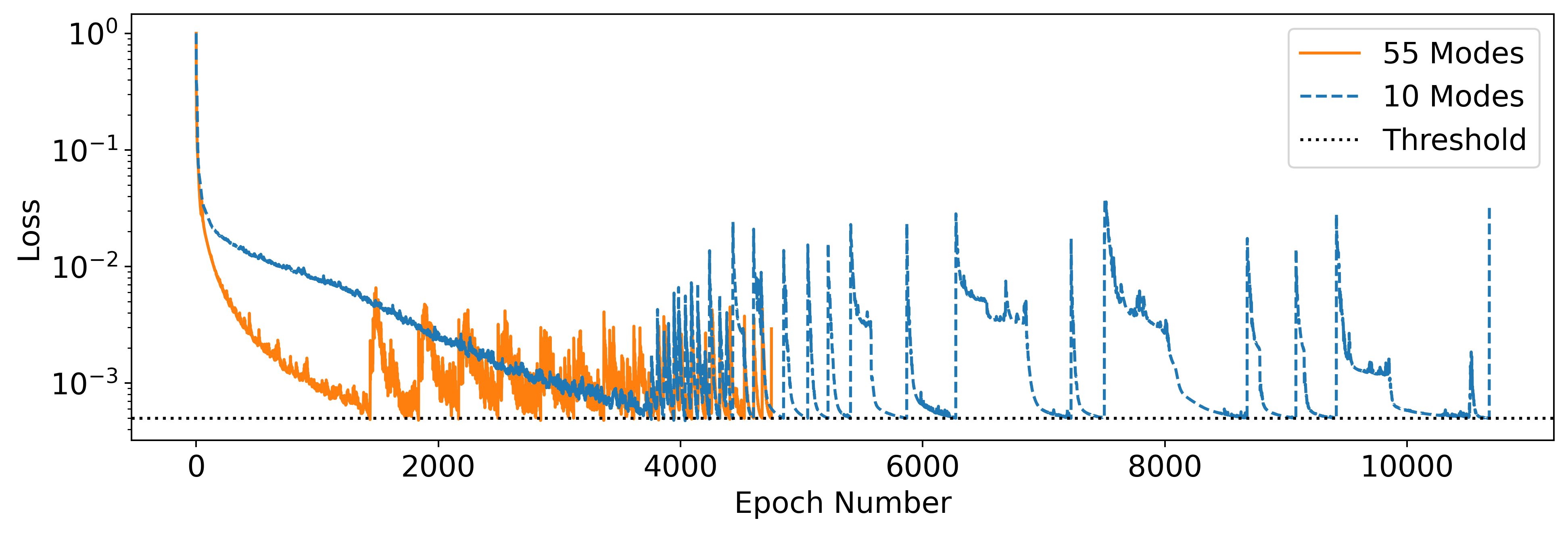}
    \caption{Loss evolution of two CD-ROM architectures using a different number of modes.}
    \label{fig:LossCurvesCompare}
\end{figure}

Figure \ref{fig:LossCurvesCompare} presents the evolution of the loss for the two models, as expected, the 55-mode model is quicker to train as it first reaches the threshold in $1400$ epochs, while it takes the 10-mode model more than twice the number of epochs to achieve the same performance. Similarly, the 55-mode model achieves the required precision on trajectories of a $1000$ time steps in only $4700$ epochs, which is again more than twice as fast as the 10-mode model.

These results confirm the interest of only using the first 10 modes to challenge our correction approach. The relatively high magnitude of the residual term to be learned, the instability embedded in the model and the low degree of resolution of the problem (only 42\% of the snapshot information) all constitute significant complexities which could arise in real world applications. While we showed that using a higher number of modes would simplify the modeling problems, this constraining choice helps demonstrate the applicability of the CD-ROM method to challenging modeling problems.

\section{Computational cost}
\label{Annex:CompCost}

While the computational cost of the overall CD-ROM approach and the way it compares to classical full order methods will strongly vary with the nature of the problem it is applied to, the simulation software used as well as the available hardware, we provide some elements of comparison with the POD Galerkin method in this appendix. Focusing on the case of the fluidic pinball (see section \ref{subsec:PinballPres}), we distinguish several components of the computational costs entailed by the CD-ROM method:

\subsection{ROM assembly}

Because most POD Galerkin models are often restricted to a very low number of modes, the cost of assembling the reduced model can often be overlooked. However, some problems might require a high number of POD modes to achieve a satisfactory resolution. For example, the fluidic pinball case requires up to a thousand modes to capture 99\% of the snapshot information, which directly impacts the cost of assembling the reduced operators $\tilde{\mathcal{L}}$ and $\tilde{\mathcal{Q}}$ in equation $\ref{eq:TensorialROM}$. Specifically, the reduced advective operator $\tilde{\mathcal{Q}}$ requires the computation of $O(n^3)$ inner products, $n$ being the number of selected POD modes.

This leads to exploding ROM assembly costs as the number of POD modes grows higher. While the assembly remains a one-time, parallelisable operation, we observed that assembling a 250 modes ROM on a $50$ cpu machine took more than a full day of computation. This underlines the interest of representing the dynamics on a low dimensional basis of modes, as assembling a thousand mode ROM would become prohibitively expensive.

\subsection{ROM Simulation}

\begin{figure}
    \centering
        \raisebox{-.5\height}{\includegraphics[width=0.65\textwidth]{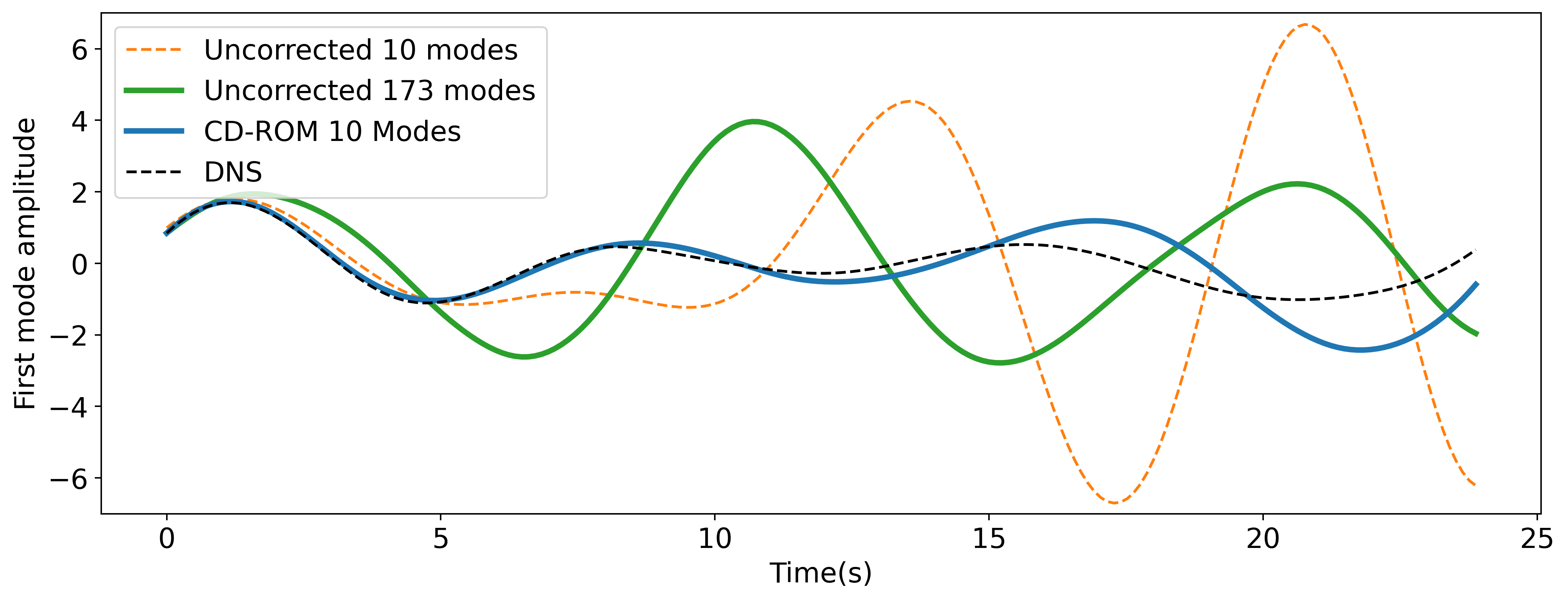}} 
        \raisebox{0\height}{\begin{tabular}{|c|r|}
        \hline
        \multicolumn{2}{|c|}{Integration times}\\
        \hline
        Uncorrected 10 modes & ~30 ms \\
        Uncorrected 173 modes & ~550 ms \\
        CD-ROM 10 modes & ~330 ms\\
        \hline
        \end{tabular}}
    \caption{Performance and computational cost of simulating different models starting from an initial condition outside the CD-ROM training basis.}
    \label{fig:ComputationalTime}
\end{figure}

Figure \ref{fig:ComputationalTime} presents the comparative simulating costs and performance of different reduced models on the fluidic pinball case. The figure shows that, although the CD-ROM does diverge from the true trajectory after some time, it performs better than its uncorrected counterparts. Specifically, the uncorrected 173 modes reduced model which captures more than $90\%$ of the snapshot information, diverges earlier than the CD-ROM, while being more expensive to simulate.

It can also be seen that the CD-ROM is significantly more expensive to simulate than the simple galerkin model. This can be explained by the cost of evaluating the neural networks embedded in the CD-ROM architecture. Indeed, neural networks require the evaluation of matrix vector products of relatively high dimension. In the pinball case, we use two hidden layers of $250$ neurons for the correction model, which explains the computational cost increase. However, it is interesting to note that the cost of evaluating a multi layer perceptron scales quadratically with its width (number of neurons per layer), while the cost of evaluating the advection term in the galerkin ROM scales cubically with the number of modes. This explains the fact that the 173 modes Galerkin ROM is more expensive to simulate than the 10 dimensional CD-ROM, while having a lower accuracy. Similar to the previous paragraph, this shows the interest of correcting a low dimensional model, rather than simply increasing the number of modes.

\subsection{Notes on Full Order Models}

Providing a detailed comparison with full order methods is outside of the scope of this study, as the computational cost of a full order model will depend on a large number of choices, ranging from simulation software implementation to numerical integration choices. We can however state that in fluid mechanics examples, the full order models were extremely expensive to simulate when compared to the reduced models studied in this paper. For example, generating the snapshot and test data for the fluidic pinball case took more than a day on a 50 cores machine. By comparison, the simulation of the trained CD-ROM for the same duration is of the order of the second on a normal computer.

\section{Hyper-Parameters \EMmodif{}{and training}}\label{sec:TrainingStrat}

This appendix details the various design and training choices made for the different models presented in the results section \ref{sec:Results}. As explained earlier in the paper, the models are trained using progressively longer prediction horizons. The main advantage of this method, as opposed to directly training with the target prediction horizon, is that potentially unstable systems in the earlier learning stages will struggle to reach long term prediction horizons without diverging, making the training extremely inefficient. Moreover, using this strategy means that a single, long DNS trajectory can be separated in numerous sub trajectories which can be batched together and simulated in parallel, yielding a very efficient training process. The procedure is presented in algorithm \ref{alg:BatchSubTraj}.

\begin{algorithm}
\caption{Batching and Sub-trajectories}
\label{alg:BatchSubTraj}
\begin{algorithmic}
\Require{$T_f,T_i$ the initial and final prediction horizons, $T_{\textit{inc}}$ the horizon increment, $\mathcal{L}_t$ the loss threshold, $f(\mbf{z,\theta})$ a CD-ROM model}
\State{$T \gets T_i$}
\State{$\mathcal{L} \gets +\infty$}
\While{$T < T_f$}
\While{$\mathcal{L} > \mathcal{L}_t$}

\State{Sample a batch of trajectories $\mbf{a}_{-\tau_{\textit{min}}\xrightarrow{}T}^\star$}

\State{Compute the initial memory $\mbf{y_0}$}

\State{$\mbf{z_0} \gets [\mbf{a_0^\star,y_0}]$}

\State{$[\mbf{a}_t,\mbf{y}_t] \gets \mbf{z_0} + \int_0^{t}f(\mbf{z;\theta}) \ddroit t$} \Comment{CD-ROM simulation}

\State{$\mathcal{L} \gets \mathcal{L}(\mbf{a}_{0\xrightarrow{}T},\mbf{a}^\star_{0\xrightarrow{}T})$}

\State{Backpropagation \& Gradient Step}

\EndWhile

\State{$T \gets T+T_{\textit{inc}}$}
\State{Reorganise batches according to the new $T$}
\EndWhile
\end{algorithmic}
\end{algorithm}

Finally, the hyper-parameters values and training details for the different models trained using the above strategy are presented below. In an effort to improve readability, the various values are organised in table \ref{tab:HyperParams}.

\begin{table}[H]
    \centering
    \begin{tabular}{|l|c c c|}
        \hline & Cylinder & Pinball & KS  \\
        \hline Memory Size & 30 & 50 & 30 \\
        Corrector Neurons & (30,30,30,30,3) & (50,250,250,250,10) & (56,150,150,150,25)\\
        Encoder Neurons	& (3,9,15,21,27) & (10,17,24,31,40)	& (26,21,16,11,5)\\
        Activation & SiLU & SiLU & SiLU\\
        Optimizer & Adam & Adam & Adam\\
        Learning Rate ($\mbf{E}$ \& $\mathcal{R}$) & $10^{-3}$ & $10^{-3}$ & $5 \times 10^{-4}$\\
        Weight Decay ($\mbf{E}$ \& $\mathcal{R}$) & $10^{-4}$ & $10^{-2}$ & $10^{-3}$\\
        Learning Rate ($\mbf{\Lambda}$) & $10^{-4}$ & $2 \times 10^{-4}$ & $5\times 10^{-4}$\\
        Weight Decay ($\mbf{\Lambda}$) & $0$ & $0$ & $0$\\
        Time Integrator & Scipy RK-45 & Scipy RK-45 & Semi Implicit 3rd order \citep{kar2006sirk3}\\
        \EMmodif{}{Training Time} & \EMmodif{}{$1/2$ day} & \EMmodif{}{1 day} & \EMmodif{}{$1/2$ day}\\ 
        \hline
    \end{tabular}
    \caption{Hyper parameters used in the training of the different models presented in section \ref{sec:Results}. The various parts of the CD-ROM are designated as follows, $\mathcal{R}$ the neural network predicting the residual of the Galerkin model, $\mbf{E}$ the memory encoder model, $\mbf{\Lambda}$ the diagonal memory matrix. \EMmodif{}{The final line presents the order of magnitude of the training times of each model on single RTX 2080 gpu.}}
    \label{tab:HyperParams}
\end{table}

\section{Regularization}
\label{sec:Regularization}
\subsection{Residual regularization}\label{sec:ResReg}

We observed that, only training the model to follow the trajectory data can lead to poor local optima with large  corrections applied to the original model. Since the magnitude of the correction is meant to be small when $r$ is sufficiently large, this tends to indicate over-fitting. This is an issue as such a model does not capture the true dynamics, and will quickly diverge when evaluated on conditions different from the training trajectory. To address this, a regularization term can be added in the model, to limit the magnitude of the corrections applied to the ROM:

\begin{equation}\label{eq:ResidualRegularization}
    J = \frac{1}{n_t} \sum_{i=1}^{n_t} \Big(\left\|\mbf{a}(i\,\Delta_t) - \mbf{a}^\star(i\,\Delta_t) \right\|^2_2 + \rho \left\|\mathcal{R}\mbf{(y}(i \, \Delta_t)\right\|^2_2 \Big)
\end{equation}
where $n_t$ is the number of time steps of length $\Delta_t$ in the optimised trajectory, $\mbf{a}^\star$ is the training trajectory data and $\rho$ is an hyper-parameter chosen to balance the importance of the regularization \textit{w.r.t.} the rest of the loss. Although this fairly simple approach already helped guiding the training, we obtained better results by adding a custom regularization loss. This loss is based on the approximate value of the residual on the true trajectory, which can be obtained by computing the time-derivative of the true reduced coordinates $\dot{\mbf{a}}^\star$ through finite differences, and computing the difference with the derivative defined by the uncorrected ROM:

\begin{equation}
    \mathcal{R}^\star = \dot{\mbf{a}}^\star - \widetilde{\mathcal{L}}\mbf{a}^\star - \mbf{a}^\star\widetilde{\mathcal{Q}}\mbf{a}^\star.
\end{equation}

With this correction data, the memory can be evaluated from the true trajectory integrating in time the following ODE:

\begin{equation}
    \dv{}{t}\mbf{y} = \mbf{E(\mbf{a}^\star;\boldsymbol{\theta}_E)} - \mbf{\Lambda y}.
\end{equation}

One can then define a regularization term $\mathcal{L}_\mathrm{corr}$ for the loss:

\begin{equation}\label{eq:TrueCorrectionLoss}
    \mathcal{L}_\mathrm{corr} = \frac{1}{n_t} \sum_{i=1}^{n_t} \left\|  \mathcal{R}(\mbf{y}(i \Delta_t)) - \dot{\mbf{a}}^\star(i \Delta_t)\right\|^2_2.
\end{equation}

The regularized loss definition was observed to lead to models with better generalization properties. Note that $\mathcal{L}_\mathrm{corr}$ can be computed for a small random subset of the a training batch at each epoch, to accelerate training while keeping a ``\textit{stochastic}'' regularization for the residual.

\subsection{Encoded space regularization}

Regularizing the encoded space can both help smooth the training process and increase the robustness of the model to unseen conditions. Taking inspiration from existing work (\cite{bollt2017edmd}), we propose to add the identity function to the encoder model. This means that useful information is embedded in memory in the form of a time convolution of the past resolved states of the system. The encoded state $\mbf{E}(\mbf{a}_t)$ is then constituted of both the reduced state $\mbf{a}_t$ and a learnable nonlinear transformation $\mathcal{MLP}(\cdot,\boldsymbol{\theta}_{\mbf{E}})$ of it:
\begin{equation}
    \mbf{E}(\mbf{a}_t,\theta_\mbf{E}) = \big[ \mbf{a}_t, \mathcal{MLP}(\mbf{a},\boldsymbol{\theta}_{\mbf{E}})]
\end{equation}
With this structure, the encoder will only be learning additional nonlinear transformations of the state, simplifying the training and introducing a level of structure in the encoded space as we ensure its first dimensions are coherent with the phase space. 

\EMmodif{}{\section{Memory Dimension}

The dimension of the memory in the CD-ROM formulation is a hyper parameter that should be chosen depending on the case and the dimension of the reduced state. Choosing an excessively low memory dimension will lead to poor prediction performance, while using too high of a dimension will negatively impact the computational cost of the corrected model and might lead to overfitting and poor numerical conditioning.

In the present examples, good results were obtained using memory dimensions ranging from $2\times$ to $10\times$ the dimension of the reduced state. However, in cases where performance becomes critical and the memory dimension should be as small as possible, a simple mask can be added to the output of the encoder to isolate necessary memory dimensions:

\begin{equation}
    \widetilde{\mathbf{E}}(x) = \mathbf{m}_\theta \odot \mathbf{E}(x)
\end{equation}

where $\odot$ is the Hadamard (pointwise) product and $\mathbf{m}_\theta \in \mathbb{R}^{n_\mathbf{E}}$ a vector with entries $m_i \in [0,1]$. By adding the $L_1$ norm ($\left\|\mathbf{m}_\theta\right\|_1$) of this vector to the training loss, the model learns a sparse mask $m_\theta$. This mask isolates redundant memory dimensions, which can then be pruned after training.}

\section*{Acknowledgements}

This work was supported by IRT SystemX, in the context of the \textit{Intelligence Artificielle et Ingénierie Augmentée} program.

\printbibliography

\end{document}